# Energy efficiency in Edge TPU vs. embedded GPU for computer-aided medical imaging segmentation and classification

José María Rodríguez Corral [a,1,*], Javier Civit-Masot [b,2], Francisco Luna-Perejón [b,3], Ignacio Díaz-Cano [a,4], Arturo Morgado-Estévez [a,5], Manuel Domínguez-Morales [b,6].

[a] Applied Robotics research group (TEP-940), School of Engineering, University of Cadiz, Avda. Universidad de Cádiz, 10, Puerto Real (Cádiz), Spain.
[b] Robotics and Technology of Computers research group (TEP-108), Architecture and Computer Technology Dept., E.T.S. Ingeniería Informática, University of Seville, Avda. Reina Mercedes, s/n, Sevilla, Spain.
[*] Corresponding author.

[1]josemaria.rodriguez@uca.es,  [2]mjavier@us.es,  [3]fluna1@us.es,  [4]ignacio.diaz@uca.es, [5]arturo.morgado@uca.es, [6]mjdominguez@us.es

**Abstract.**

In this work, we evaluate the energy usage of fully embedded medical diagnosis aids based on both segmentation and classification of medical images implemented on Edge TPU and embedded GPU processors. We use glaucoma diagnosis based on color fundus images as an example to show the possibility of performing segmentation and classification in real time on embedded boards and to highlight the different energy requirements of the studied implementations.

Several other works develop the use of segmentation and feature extraction techniques to detect glaucoma, among many other pathologies, with deep neural networks. Memory limitations and low processing capabilities of embedded accelerated systems (EAS) limit their use for deep network-based system training. However, including specific acceleration hardware, such as NVIDIA's Maxwell GPU or Google's Edge TPU, enables them to perform inferences using complex pre-trained networks in very reasonable times.

In this study, we evaluate the timing and energy performance of two EAS equipped with Machine Learning (ML) accelerators executing an example diagnostic tool developed in a previous work. For optic disc (OD) and cup (OC) segmentation, the obtained prediction times per image are under 29 and 43 ms using Edge TPUs and Maxwell GPUs, respectively. Prediction times for the classification subsystem are lower than 10 and 14 ms for Edge TPUs and Maxwell GPUs, respectively. Regarding energy usage, in approximate terms, for OD segmentation Edge TPUs and Maxwell GPUs use 38 and 190 mJ per image, respectively. For fundus classification, Edge TPUs and Maxwell GPUs use 45 and 70 mJ, respectively.

# 1. Introduction.

There are multiple causes that affect the time to obtain a medical diagnosis. Among them are the lack of medical professionals and the difficulty of access to healthcare services. In this vein, the World Health Organization (WHO) estimates that by 2030, there will be a global shortage of about 14 million health professionals (WHO, 2006). This lack of health personnel can lead to an overload of healthcare systems, resulting in increased waiting times for care, a reduced quality of health services and, ultimately, the collapse of the health system (Golz et al., 2022).

Moreover, the lack of access to healthcare is a serious problem that affects many people around the world, especially those who live in Third World countries, rural areas, or have disabilities or reduced mobility. According to the WHO, more than half of the world's population does not have access to essential health services (WHO, 2020). In low-income countries, lack of access to healthcare is one of the leading causes of mortality. As indicated in WHO (2020), 45% of child deaths worldwide occur in low-income countries due to lack of access to basic health services. In addition, many people living in rural areas do not have access to adequate healthcare services due to lack of infrastructure and trained medical personnel. Patients in rural areas must travel far from their homes to assess their health status, which makes access to health services difficult and often impossible (Sanz-Tolosana, Oliva-Serrano, 2021).

Moreover, people with disabilities or reduced mobility also face many challenges in accessing adequate health services. According to WHO, more than one billion people worldwide have some form of disability and are more likely to experience barriers in accessing health care than people without disabilities (WHO, 2021). Lack of physical accessibility and lack of trained personnel to care for people with disabilities are some of the main challenges faced by these people.

These problems, among others, have led to the proliferation in recent years of diagnostic support systems that reduce long hospital queues and help reduce the time it takes for a healthcare professional to provide a diagnosis. In this sense, Machine Learning based systems have been widely used to design classifiers that provide a fast response to system input information (physiological, basal or even image information).

Artificial Intelligence (AI) and Machine Learning (ML) are present in many devices and services used in our everyday life, Health care related devices are not an exception, since the application of Deep Learning techniques to the medical image analysis has significantly increased in recent years (Litjens et al., 2017; Teikari et al., 2019; Chen et al., 2020; Prabhu and Verma, 2021). Two important Deep Learning uses in medical imaging are classification and segmentation.

In image classification (Litjens et al., 2017), one or several images are used as input, and a small set of diagnostic variable values - for example, disease present or not, or a numeric level of disease severity - are provided as output. Image classification can be performed, among other possible alternatives, by using Convolutional Neural Networks (Hesamian et al., 2019; Chen et al., 2020; Chollet, 2021).

On the other hand, segmentation is related to the detection of object boundaries within an image, performed either automatically or semi-automatically. In this sense, the segmentation of substructures allows, for example, the study of parameters related to shape and volume in brain (Niepceron et al., 2020) or cardiac analysis (Chen et al., 2020), for example.

Many networks with widely different structures have been used for medical image segmentation (Hesamian et al., 2019; Litjens et al., 2017). An U-Net (Fig. 1) is a specific type of fully convolutional network (FCN) that was initially proposed by Ronneberger et al. (2015) and has been used with different medical image types, such as Computerized Tomography (CT), Magnetic Resonance Imaging (MRI), X-Ray and Ultrasound (US).

However, to extend the use of these AI-based diagnostic support systems, the processing needs to be done on a stand-alone device so that, in the case of rural areas or third world countries, the equipment can be easily relocated. This is a problem when working with images, as it requires a high computational cost (even more so if the segmentation process is included).

In this vein, single-board computers (SBCs) are small low-cost computers where processor and memory are integrated into a system on chip (SoC). These devices are very widely used and allow developers to interact with sensors and actuators in applications that include computing, IoT (Hassan et al., 2017) and robotics among others (Ariza and Baez, 2021). At present, some SBCs, that we call embedded accelerated systems (EAS), are equipped with ML hardware, which makes them ideal targets for Edge computing based Artificial Intelligence (EdgeAI) applications (Baller et al., 2021). Some of these EAS are Coral Dev Board (Google LLC, 2020a, 2020b) and Jetson Nano 2GB (NVIDIA Corp., 2021a).

Excluding more powerful models of the Jetson series (e.g., NVIDIA Jetson AGX Xavier and Jetson AGX Orin Developer Kits), it is technically inviable to train relatively complex deep neural networks using EAS, because of their constrained resources in terms of main memory size and processing capacity. However, ML accelerators enables them to perform real time inferencing using complex pre-trained networks, as inferencing is computationally a much less demanding task.

Some of these platforms are designed to deal with AI-based classification and segmentation problems, but few papers make use of them in real-world problems (and even fewer evaluate their performance).

Considering now our example problem, glaucoma is a retinal disease that encompasses a set of progressive neuropathies that causes damage to the optic nerve head in the back of the eye, leading to loss of the visual field and, finally, blindness (Weinreb et al., 2014). The exam of the optic nerve head, where cell axons leave the eye forming the optic disc (Bourne, 2006; Diaz-Pinto et al., 2019), is critical for to the detection of this disease. In retina fundus images, the optic disc (Fig. 2) can be divided into two zones: a central one known as the optic cup and a peripheral part, called neuro-retinal rim. This is a region consisting mainly of nerve fibers.

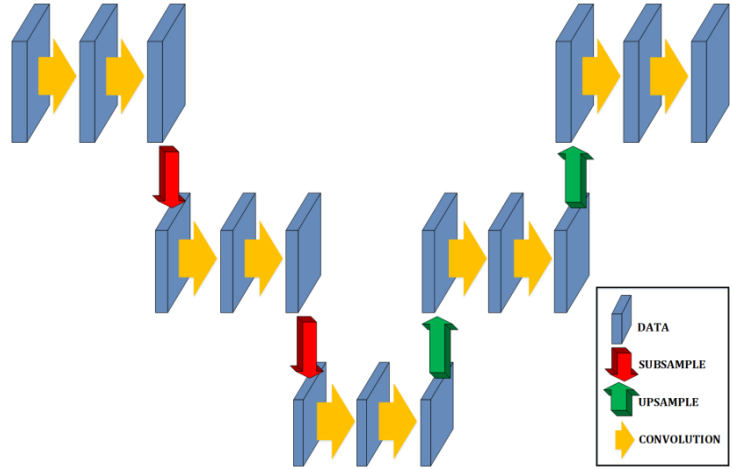

Fig. 1. Basic three-layer U-Net.

Ophthalmologists perform Glaucoma diagnosis using, among other data, several indicators obtained from fundus images. A widely used metric is the cup-to disc-ratio (Patel and Patel, 2018; Zulfira et al., 2021). The CDR is defined as the relation between the optic cup (OC) and the optic disc (OD) diameters. Another widely used criteria for glaucoma assessment is the ISNT rule (Bourne, 2006; Das et al., 2016) which is based on the shape of the neuro-retinal rim. Correct segmentation of both OC and OD is essential to calculate the CDR (Barros et al., 2020; Patel and Patel, 2018) and also for the application of the ISNT rule (Barros et al., 2020; Nath and Dandapat, 2012).

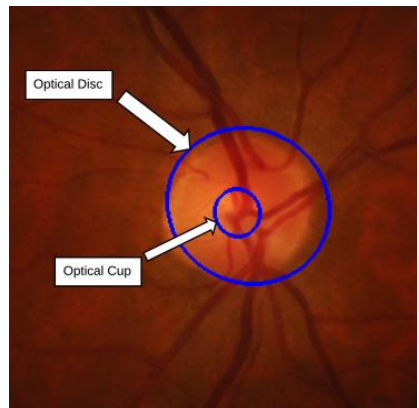

Fig. 2. OD and OC in eye fundus image.

In this vein, this work aims to evaluate and compare the performance of two EAS based on completely different approaches to ML hardware acceleration: Coral Dev Board and Jetson Nano 2GB Developer Kit. These widely different systems are both used to implement, as a good example of the tasks needed for an image-based diagnosis tool, the generalized U-Net models for segmentation of OD and OC, and the MobileNet V2 based-model for fundus classification proposed in Civit-Masot et al. (2020, 2021).

More specifically, the aim of this study is not just to prove the practical viability of embedding these models in the two EAS, but also testing if they can perform inferences in reasonable times. In this regard, we also want to compare the inference times provided by both EAS with those obtained on non-embedded cloud-based GPUs and TPUs.

The use in everyday medical practice of EAS to perform medical image classification and segmentation tasks in assumable times is very convenient, since ML accelerators can be integrated in a broad range of lightweight medical instruments to perform segmentation and classification tasks autonomously, without requiring an external PC. Moreover, the use of EAS for medical image analysis protects the privacy of patients' health data, since their private information is used locally by the embedded systems instead of being processed by remote servers. Finally, since Cloud GPUs and TPUs are usually free for non-commercial uses only, the use of ML accelerators, such as the GPUs embedded in NVIDIA Jetson boards or the Google Edge TPU used by the Coral Dev Board, are interesting alternatives to consider.

In this sense, it is important to measure the energy efficiency of ML accelerators when performing inferences as they can be embedded in devices that operate autonomously using batteries. Therefore, in this work we measure the energy usage of the two EAS - Coral Dev Board and Jetson Nano 2GB - when performing OD and OC segmentation as well as fundus classification. The obtained results are shown and discussed in the corresponding sections.

The rest of the document is organized in the following way: Section 2 presents the background and a set of related works. Section 3 describes the methodological aspects. The obtained results are presented and discussed in Sections 4 and 5, respectively. The conclusions of this study are presented in Section 6, and several future lines of work are proposed.

**2. Background.**

In Civit-Masot et al. (2020), a diagnostic aid tool to detect glaucoma using fundus images built around a segmentation and a classification subsystem was proposed. These two subsystems were independently trained and tested, and their results were combined to improve glaucoma diagnosis. The first subsystem applies ML techniques to segment optic disc and cup independently and extract their positional and physical features. The second one applies transfer learning techniques to a pre-trained CNN to detect glaucoma through the analysis of the complete fundus images. The results showed that this system achieves a higher classification rate than those used in previous works. A practical implementation of the segmentation subsystem on an Edge TPU device was discussed in Civit-Masot et al. (2021).

As for the use of EAS for medical image classification, in Rehman et al., (2020) the authors developed an image processing method for melanoma skin lesion detection at the initial stage using an NVIDIA Jetson Nano Developer Kit, consisting of five phases: load of a skin lesion dermoscopic image on the device, grayscale conversion of the color image, segmentation of the region of interest, noise removal and classification. In the last phase, the lesion features were extracted and classified into three categories: benign, suspicious, and malignant. The experiments were performed on PH2 and ISIC publicly available skin lesion datasets. In comparison with other published works on skin cancer identification, the proposed approach provided better results, such as average accuracies of 93.5% and 91.45% for PH2 and ISIC datasets, respectively.

Moreover, the work presented in Prabhu and Verma, (2021) proposes a model to classify a healthy skin and one affected by Diabetic Foot Ulcer (DFU) from plantar thermograms using Deep Learning algorithms. The implementation of the model on a Jetson Nano Developer Kit shows the suitability of the solution for deployment on embedded devices. The implementation performance was compared with that of existing CNNs. As a result, the trained model (a modified version of DenseNet) achieved a maximum accuracy of 97.9%, making it suitable for automatic DFU classification in order to aid clinicians in their diagnosis.

More specifically, in relation to the use of mobile devices, single-board computers and embedded systems to perform segmentation and classification of fundus images, Martins et al., (2020) propose an interpretable computer-aided diagnosis (CAD) pipeline to diagnose glaucoma from fundus images that executes in mobile devices. The authors mixed various publicly available datasets and used them to train CNNs in order to perform classifications and segmentations. These segmentation and classification models were then used to make a pipeline that, in addition to the segmentation of key structures and the calculation of various morphological features, showed a glaucoma confidence level, resulting in an interpretable CAD in a similar way to that implemented in the cloud in Civit-Masot et al. (2020). To test the performance of the developed system in a resource constrained device, the processing chain was tested in a smartphone (Samsung Galaxy S8) using a mobile application developed for this purpose, where memory requirements as well as CPU and GPU execution times were evaluated. Comparable or slightly better metrics than current CAD systems for glaucoma assessment were achieved.

Also, a Deep Learning approach to assess the quality of eye fundus images was presented in Pérez et al. (2020). The proposed model (MFQ-Net) is sufficiently small to be deployed in a smartphone and was validated using a public fundus dataset with two sets of annotations providing good accuracy results: 0.911 in the binary classification task (accepted and rejected), and 0.856 in a three-class classification task (good, usable, and bad). The authors also measured the memory requirements and the classification average time for the binary (MFQ-Net Binary) and the three-class (MFQ-Net Three-class) models on an Android 9.0 smartphone. The number of parameters of the eye fundus quality classification system proposed by the authors is smaller compared with other state-of-the-art models, making this approach interesting for execution on a mobile device.

Moreover, a methodology is proposed in Washburn et al., (2021) for hardware-based detection of exudates (i.e., proteins and lipids filtered from abnormal blood vessels), which are one of the main signs that identifies the diabetic retinopathy, along with micro-aneurysms and hemorrhages. In the proposed method, retinal images were selected from the DRIVE and STARE public databases. After being preprocessed, the images were segmented using the Fuzzy C-Means Clustering segmentation algorithm. This algorithm was implemented on a Jetson Nano Developer Kit using the RAPIDS CUDA Machine Learning Libraries, and executed completely on its GPU, achieving a sensitivity of 98.3% and a specificity of 98.9%. Finally, the results showed an improvement in terms of execution time and power consumption compared to other implementations.

On the other hand, in relation to the performance evaluation of EAS, and how deep neural networks performs on these devices, in Baller et al., (2021) the efficiency in terms of power consumption and inference time of four SBCs - Raspberry Pi 4, NVIDIA Jetson

Nano, Google Coral Dev Board and Asus Tinker Edge R - and the Arduino Nano 33 BLE microcontroller, was compared on different deep learning frameworks - TensorFlow, TensorRT, TensorFlow Lite and RKNN-Toolkit - and models such as MobileNet V1 and MobileNet V2. The authors also provided a method to measure accuracy, inference time and power consumption for the devices used, that can be easily extended to other architectures. The results showed that, for TensorFlow-based quantized models, Coral Dev Board provided the best performance for power consumption and Edge TPU inference time. However, considering that Jetson Nano does not include specific ML hardware to perform inferences using quantized models, since it integrates a Maxwell GPU instead of an Edge TPU, this EAS outperformed the other devices when performing GPU inferences with the non-quantized version of MobileNet V2 using the TensorRT framework.

Finally, Table 1 is included as a summary of this section.

| Study reference | IA models used | Type of process unit/computer/device | Performance metrics | Results obtained | Problem/challenge addressed |
|---|---|---|---|---|---|
| Civit-Masot et al. (2020) | U-Net | GPU and TPU (Google services) | Dice coefficient | 0.93 (GPU/TPU) | Fundus image segmentation and classification for Glaucoma diagnosis support |
| Civit-Masot et al. (2021) | U-Net | CPU, GPU and TPU (Google services) Edge TPU (Coral Dev Board) | Dice coefficient | 0.91 (CPU/GPU/TPU) 0.90 (Edge TPU) | Fundus image segmentation for Glaucoma diagnosis support and efficiency comparison |
| | | | Inference time | Best performance: Edge TPU (Coral Dev Board) | |
| Rehman et al. (2020) | ABCD rule and Total Dermoscopy Score (TDS) | NVIDIA Jetson Nano Developer Kit | Accuracy | 93.5% (PH2 dataset), 91.45% (ISIC dataset) | Melanoma skin lesion detection |
| Prabhu and Verma, (2021) | DenseNet | Jetson Nano Developer Kit | Accuracy | 97.9% | Diabetic Foot Ulcer (DFU) classification |
| Martins et al., (2020) | Custom CNNs | CPU and GPU in mobile devices | Accuracy Sensitivity AUC | 87% 85% 93% | Glaucoma diagnosis from fundus images |
| Pérez et al. (2020) | MFQ-Net (Deep Learning model) | Mobile device | Accuracy | 91.1% (Binary), 85.6% (Three-class) | Quality assessment of fundus images |
| Washburn et al., (2021) | Fuzzy C-Means Clustering algorithm | Jetson Nano Developer Kit | Sensitivity Specificity | 98.3% 98.9% | Exudate detection to identify diabetic retinopathy |
| Baller et al., (2021) | Various deep learning frameworks | Asus Tinker Edge R, Coral Dev Board, Jetson Nano, Raspberry Pi 4, Arduino Nano 33 BLE | Accuracy | 92.5% | Efficiency (power consumption and inference time) comparison of four SBCs and one microcontroller for deep learning model execution |
| | | | Inference time and power consumption | Best performance: 1. Coral Dev Board (Edge TPU inference using quantized models) 2. Jetson Nano (GPU inference using non-quantized MobileNet V2) | |

Table 1. Contributions.

## 3. Materials and methods.

In this section we will briefly describe the hardware resources, the datasets used and the classification network implementation. Finally, we will specify the parameters of the chosen generalized U-Nets, and we will outline the methodology used to perform the experimental tests.

3.1. Hardware.

The Jetson Nano 2GB Developer Kit (NVIDIA Corp., 2021a) is equipped with a 128-core NVIDIA Maxwell architecture-based GPU for delivering high AI performance at a low price, a Quad-core ARM A57 @ 1.43 GHz and 2 GB 64-bit LPDDR4 SDRAM (Fig. 3).

The board provides various USB 2.0/3.0 ports, a Gigabit Ethernet port, an HDMI display interface, a 40 I/O pin header and a CSI-2 camera interface among other connectors and uses a microSD card as storage system. This kit is intended for introducing anyone interested in learning embedded AI fundamentals. It is supported by a series of comprehensive tutorials and an active developer community that make ready-to-build open-source projects.

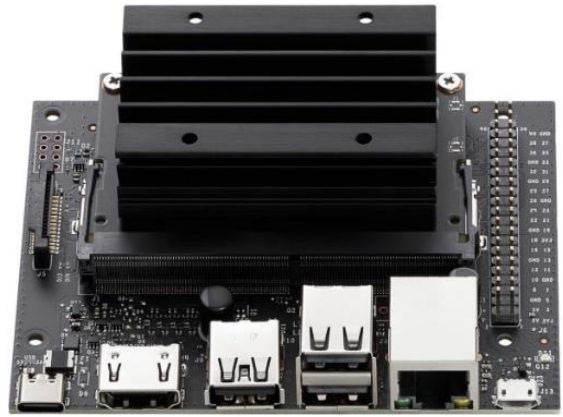

Fig. 3. Jetson Nano 2GB.

NVIDIA JetPack SDK (NVIDIA Corp., 2021b) bundles Jetson platform software including TensorRT (NVIDIA Corp., 2021c, 2021d), cuDNN (NVIDIA Corp., 2021e) and CUDA Toolkit (NVIDIA Corp., 2021f) among other tools, all built on top of L4T - a Linux based system software distribution by NVIDIA for the Tegra processor series - with LTS Linux kernel (NVIDIA Corp., 2021g).

The Coral Dev Board (Google LLC, 2020a, 2020b) includes a small ASIC (Google TPU coprocessor) which delivers high performance ML inferencing for TensorFlow Lite models (Google LLC, 2020c). The board (Fig. 4) is equipped with a System-on-module (SoM), an integrated system that can be included in a custom board for purposes of production and can be bought separately.

The SoM incorporates a NXP's IMX 8M system on chip, 1 or 4 GB LPDDR4 SDRAM, 8 GB eMMC memory, a Google Edge TPU coprocessor and Bluetooth 4.2 and Wi-Fi 802.11a/b/g/n/ac connection capabilities. The SoC integrates a Quad-core ARM Cortex-A53 @ 1.5 GHz processor, an Arm Cortex-M4F processor and a Vivante GC7000Lite GPU.

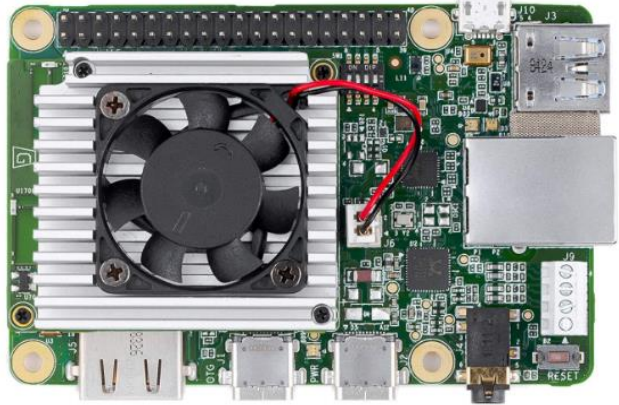

Fig. 4. Coral Dev Board.

This EAS runs a lightweight operating system based on Debian Linux named Mendel. The board provides various USB 2.0/3.0 ports, a Gigabit Ethernet port, a 40 I/O pin header, a DSI display interface and a CSI-2 camera interface.

Finally, Table 2 is shown as a conclusion of this subsection.

|  | **Jetson Nano 2GB** | **Coral Dev Board** |
| --- | --- | --- |
| Memory size | 2 GB 64-bit LPDDR4 SDRAM | 1 or 4 GB LPDDR4 SDRAM |
| CPU | Quad-core ARM A57 @ 1.43 GHz | Quad-core ARM Cortex-A53 @ 1.5 GHz and Arm Cortex-M4F |
| GPU | 128-core NVIDIA Maxwell architecture-based GPU | Vivante GC7000Lite |
| TPU | - | Google TPU coprocessor |
| Connectivity | USB 2.0/3.0 ports, Gigabit Ethernet port, HDMI display, CSI-2 camera interface | USB 2.0/3.0 ports, Gigabit Ethernet port, HDMI display, CSI-2 camera interface, DSI display interface, Bluetooth 4.2, Wi-Fi 802.11a/b/g/n/ac |
| GPIO | 40 I/O pin header | 40 I/O pin header |
| Storage system | microSD card | 8 GB eMMC |
| Operating System | L4T (Linux based system) | Mendel (Debian Linux based system) |
| Software | TensorRT, cuDNN, CUDA Toolkit | TensorFlow Lite |

Table 2. Comparison of features between Jetson Nano 2GB and Coral Dev Board.

3.2. Datasets.

In this work, we use the same publicly available and very widely used eye fundus datasets as in Civit-Masot et al. (2020, 2021): DRISHTI-GS (Sivaswamy et al., 2014) and RIM-ONE-v3 (Fumero et al., 2011). Both provide human expert segmentation data for OD and OC, as well as labels that indicate if the images come from glaucoma patients or healthy ones.

From the eye fundus images, several augmented datasets with different sizes have been generated to be used in the experimental tests (section 4) and thus obtain the corresponding prediction times.

3.3. Network architecture.

In this subsection, we will address those aspects relating to the structure of the system and the design of the experimental tests. Regarding the U-Nets, we have selected the same ones used in Civit-Masot et al. (2021), since their sizes allow an embedded implementation. First, a U-Net with 6 levels, 40 filters in the initial layer, a layer-to-layer increment ratio (IR) of 1.1 and 0.9 million trainable parameters (MTP) has been chosen for segmentation of OD. And a U-Net with 6 levels, 64 filters in the initial layer, an IR of 1.1 and 2.4 MTP has been chosen for segmentation of OC.

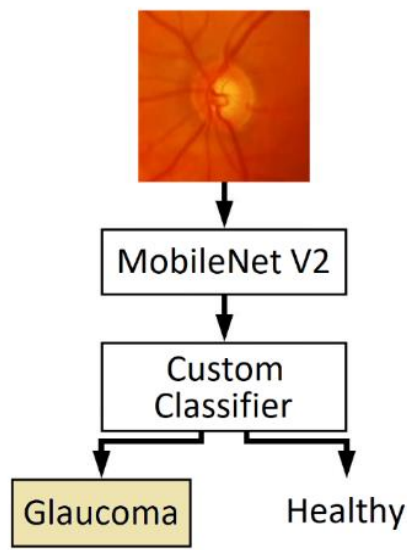

Fig. 5. Classification subsystem.

The classification network (Civit-Masot et al., 2020), is based in MobileNet V2 (Sandler et al., 2018). This network is light (less than 2.5M parameters) and thus an embedded implementation is feasible. Its accuracy on the ImageNet challenge (Russakovsky et al., 2015) is very similar to that of VGG16 (Simonyan and Zisserman, 2015), a network successfully used by other research (Diaz-Pinto et al., 2019) for fundus image classification.

However, that network is significantly larger (about 15M parameters) and makes an embedded implementation more difficult. Finally, the MobileNet V2 accuracy density - the accuracy divided by the number of parameters - is an order of magnitude higher than that of VGG16 (Bianco et al., 2018).

In order to implement the classification subsystem (Fig. 5), the top layers of the original MobileNet V2 were removed and replaced by an average polling layer whose output is flattened and sent to a dense layer with 80 nodes, a dropout stage and a final two-node dense layer to discriminate between the required classes ("glaucoma" or "healthy").

Fig. 6 presents the implementation of the full system. Finally, we provide a brief description of the methodology used to perform the experimental tests:

- Load of image dataset.
- Definition and compilation of TensorFlow model, and load of weights to perform the experimental tests using the iPython notebooks.
- Export of TensorFlow model to Open Neural Network Exchange format (ONNX), generation of CUDA inference engine from the ONNX model (Abbasian et al., 2021; NVIDIA Corp., 2021c), and load of the engine to perform the experimental tests using the Jetson Nano 2GB EAS.
- Conversion and load of TensorFlow Lite model to perform the experimental tests using the Coral Dev Board.
- Execution of inferences with time measurement.

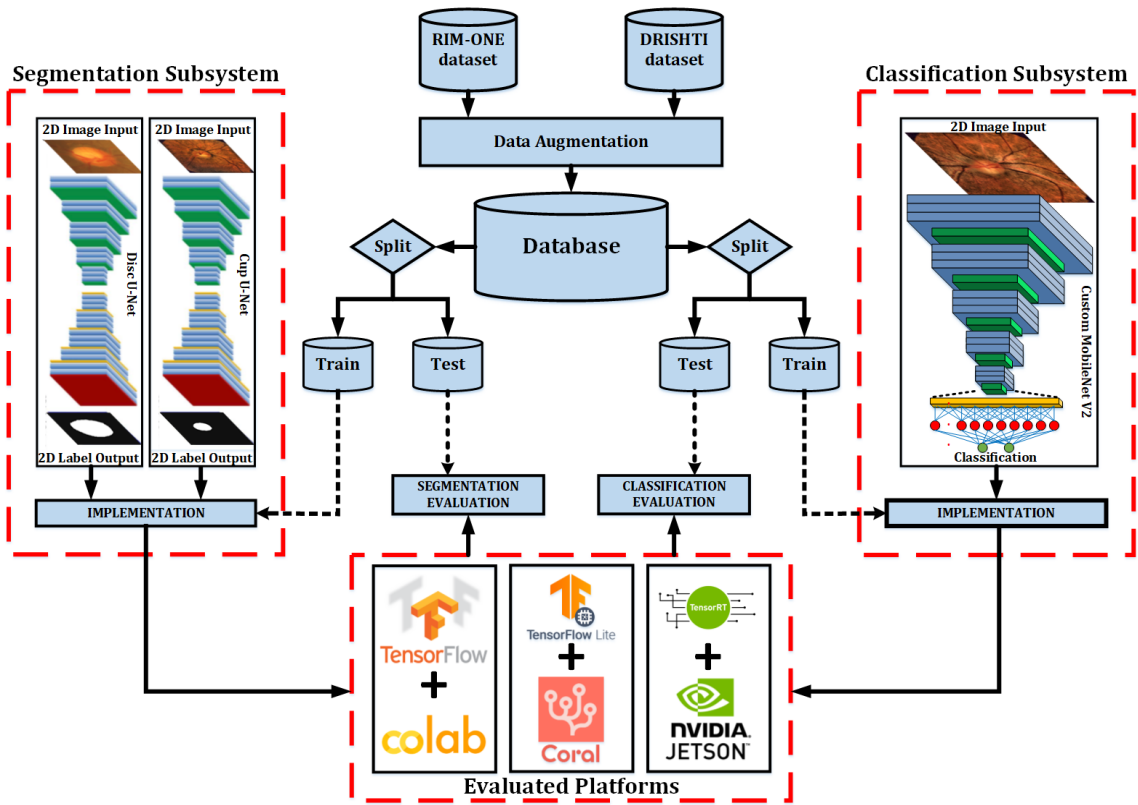

Fig. 6. Full system implementation.

## 4. Results.

In this section, the inferences performed by the OD and OC models and the fundus classification model, implemented in Google Cloud, will be compared with those ones generated by Coral Dev Board and Jetson Nano 2GB EAS in order to prove that there is no significant difference in the quality of the predictions and, consequently, both EAS are suitable for segmentation and classification of fundus images. Then, a set of results regarding inference times will be obtained to prove that both EAS perform segmentation of OD and OC and fundus classification in reasonable times (i.e., a few milliseconds).

Finally. the consumption values for the two EAS when performing inferences will be obtained in order to study the different energy requirements of the proposed implementations.

Firstly, as a performance measure of the proposed segmentation system, the Dice coefficients achieved by our implementation in comparison with other Deep Learning based options using the same datasets (subsection 3.2). are shown in Table 3 The Dice coefficient is defined, as usual, as twice the number of active pixels in the intersection of the true and the predicted masks divided by the sum of the active pixels in both masks (Sørensen, 1948).

These results are comparable with other research works, as we can see. While in Civit-Masot et al. (2020) the same U-Net was used to segment the OD and OC, in this study, as in Civit-Masot et al. (2021), smaller networks have been used for OD and OC segmentation, which are suitable for embedded implementation. Doing so decreases the number of trainable parameters and obviously has a certain impact on the segmentation performance.

| Author | Disc DRISHTI | Disc RIM-ONE | Cup DRISHTI | Cup RIM-ONE |
|---|---|---|---|---|
| Sevastopolsky (2017) | - | 0.94 | - | 0.82 |
| Shankaranarayana et al. (2017) | - | 0.98 | - | 0.94 |
| Zilly et al. (2017) | 0.97 | - | 0.87 | - |
| Al-Bander et al. (2018) | 0.95 | 0.90 | 0.83 | 0.69 |
| Civit-Masot et al. (2020) | 0.93 | 0.92 | 0.89 | 0.84 |
| Cloud GPU/TPU (this work) | 0.91 | 0.86 | 0.87 | 0.81 |

Table 3. Dice coefficients for segmentation of OD and OC.

In order to obtain a measure of the difference between the predictions performed by the optic disc and cup models implemented in Google Cloud using GPU and TPU, and those ones generated by Coral Dev Board and Jetson Nano 2GB EAS, the Dice coefficient has also been used as a comparison criterion.

Table 4 shows the mean values of the Dice coefficient ratios along with their standard deviations obtained from the comparison image by image between the predictions performed by the segmentation models on the combination of DRISHTI-GS and RIM-ONE datasets using the resources of Google Cloud, and Coral Dev Board and Jetson Nano 2GB EAS. For each comparison, the images corresponding to predictions performed by the specific U-Net - OD or OC - using the resources of Google Cloud - GPU and TPU - and TensorFlow are taken as reference values.

|  | Google Colab and Jetson Nano 2GB | Google Colab and Coral Dev Board |
|---|---|---|
| Optic disc | 0.999 ± 0.001 | 0.976 ± 0.024 |
| Optic cup | 0.999 ± 0.001 | 0.969 ± 0.029 |

Table 4. Dice coefficient ratios for predictions (images).

From the results obtained, we can conclude that the images obtained from the predictions performed by Google Cloud resources are very similar to the ones generated by Jetson Nano 2GB. This result is logical considering that CUDA inference engines generated for optic disc and cup segmentation use the same 32-bit floating point precision (NVIDIA Corp., 2021d) as used in the cloud implementation.

However, the values for the Dice coefficient ratios obtained from the comparisons between the images generated by the inferences performed by Google Cloud resources and by Coral Dev Board are a bit smaller than the previous ones, but not excessively. In fact, the mean values for optic disc and cup remain close to one. This result can be explained by the use of 8-bit integer precision in the quantized models used by Coral Dev Board (Google, 2021b; Google LLC, 2020d, 2020e).

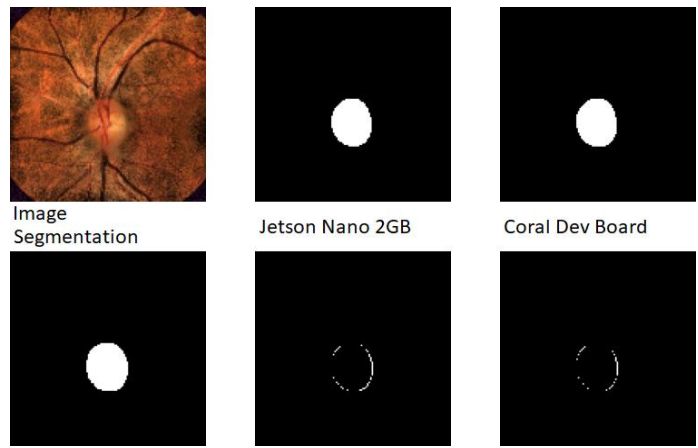

Fig. 7. Images of OD, segmentation, and inferences performed by Jetson Nano 2GB and Coral Dev Board.

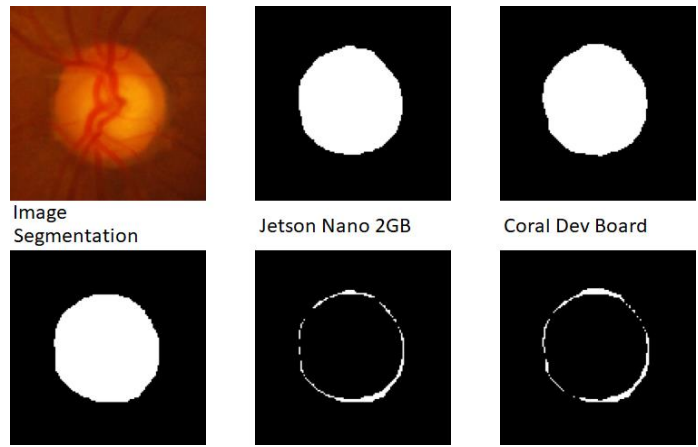

Fig. 8. Images of OC, segmentation, and inferences performed by Jetson Nano 2GB and Coral Dev Board.

Figs. 7 and 8 show images of the OD and OC respectively along with the segmentations performed by human experts, the inferences made by Jetson Nano 2GB and Coral Dev Board EAS, and the differences between each segmentation and the corresponding predictions.

Concerning the classification network, since the CNN is the same one proposed in Civit-Masot et al. (2020), the normalized confusion matrices from that work for segmentation by expert ophthalmologists and for the classifier based on MobileNet V2 (Tables 5 and 6) are included here only for completeness.

|  | Glaucoma | Healthy |
|---|---|---|
| Glaucoma | 0.67 | 0.33 |
| Healthy | 0.13 | 0.87 |

Table 5. Confusion matrix based on human segmentation.

|  | Glaucoma | Healthy |
|---|---|---|
| Glaucoma | 0.81 | 0.19 |
| Healthy | 0.17 | 0.83 |

Table 6. Confusion matrix based on MobileNet V2.

To compare the predictions performed by the classification model implemented in Google Cloud using GPU and TPU with those ones generated by Coral Dev Board and Jetson Nano 2GB EAS, the mean error has been used as a measure of similarity. Since there are two probability results for each prediction - one for glaucomatous eye class and one for healthy eye class - the error for a specific prediction is taken as the maximum of the two differences in absolute values.

For each difference, the probability results corresponding to the prediction performed by the classification network using Google Cloud resources and TensorFlow are considered as the reference values, whereas the second term of the comparison is the result of the classification performed by Jetson Nano 2GB or Coral Dev Board. Table 7 shows the mean errors calculated as the sum of the differences - in absolute values - for the predictions on the combination of the DRISHTI-GS and RIM-ONE datasets, divided by the total number of predictions (i.e., the size of the combined datasets).

|  | Google Colab and Jetson Nano 2GB | Google Colab and Coral Dev Board |
|---|---|---|
| Eye fundus image | $0.001 \pm 0.001$ | $0.046 \pm 0.089$ |

Table 7. Mean errors for predictions (classifications).

As with the Dice coefficients calculated to compare the predictions for the optic disc and cup, the predictions performed by Google Cloud resources are very similar to the ones generated by Jetson Nano 2GB, since the CUDA inference engine for eye fundus classification also uses 32-bit floating point precision (NVIDIA Corp., 2021d).

However, the value for the mean error obtained from the comparisons between the classifications (i.e., probability results) performed by Google Cloud resources and by Coral Dev Board is greater than the previous one but remain close to zero. Again, the 8-bit integer precision of the quantized models used by Coral Dev Board (Google, 2021b; Google LLC, 2020d, 2020e) clearly affects the accuracy of predictions.

In practice, these errors do not change the predicted result for any image in both EAS.

In the rest of the document, we will only show results regarding inference times and energy efficiency since, after showing that there is no significant difference in the quality of the predictions, this is the primary aim of this study. Firstly, a set of times has been obtained using the two U-Nets implemented for segmentation of OD and OC, as well as using the CNN implemented for eye fundus classification. These results have been obtained for Google Cloud GPU and TPU, and will be considered as reference times, so that we can compare them with inference times obtained for Jetson Nano 2GB and Coral Dev Board EAS.

A TPU device has four chips and two TPU cores in each chip (Google, 2021a). A TPU core consists of one Matrix Multiply Unit (MXU), a Vector Processing Unit (VPU) and a Scalar Unit. The VPU is used for general computations, and the scalar unit for control flow and calculating memory addresses among other operations. The virtual machines where Colab notebooks execute communicate with Cloud TPUs over a gRPC[1] network. To perform inferences using Cloud GPUs, an NVIDIA Tesla K80 has been used.

As indicated in subsection 3.2, from the images of RIM-ONE-v3 and DRISHTI-GS datasets, several new datasets with different sizes have been generated to be used in the experimental tests and thus obtain the corresponding prediction times.

In order to facilitate the comprehension of the text, the obtained results have been included in Appendix A, and have been organized in three tables: Image prediction times for segmentation of optic disc (Table A.1) and optic cup (Table A.2), and image prediction times for fundus classification (Table A.3). Also, graphical representations from the image prediction times are shown later in this section (figs. 9, 10 and 11).

The first two columns of results of Tables A.1 and A.2 are the prediction times per image obtained using the Colaboratory iPython notebook development environment[2] for segmentation of OD and OC respectively, whereas the first two columns of results of Table A.3 show the prediction times per image for fundus classification. 2.7.0 version of TensorFlow and Keras (Chollet, 2021) has been used. Given that TensorFlow performs predictions on complete datasets, the inference time per image has been calculated as the prediction time for a dataset divided by its number of elements.

The first prediction on a dataset using Cloud TPU also involves sending the model data through the network to copy them into the TPU memory. Moreover, when using Cloud GPU, the first prediction also takes more time due to the need to perform memory initializations and allocations. Therefore, for each dataset, the image prediction time has been calculated from the next ten predictions on the dataset after discarding the initial one. Thus, the mean prediction time per image and the standard deviation are shown in the two first columns of results of Tables A.1, A.2 and A.3.

After obtaining the first set of results, the next step is to perform predictions and measure the times using the Coral Dev Board and Jetson Nano 2GB EAS. As for Coral Dev Board, the procedure to follow consists of quantizing the Keras models (Google, 2021b) and obtaining the prediction times using the Edge TPU (Google LLC, 2020d, 2020f) for optic disc and cup segmentation, and eye fundus classification. The results obtained are presented in the third column of results of Tables A.1, A.2 and A.3.

---

[1] https://grpc.io/
[2] https://colab.research.google.com

Since TensorFlow Lite is used to perform inferences on the Edge TPU (Google LLC, 2020d), the post-training quantization technique (Google, 2021b; Google LLC, 2020e) has been applied to adapt the original TensorFlow models for the CNN based on MobileNet V2 and for the two U-Nets to the TensorFlow Lite format. The framework version used in this work is 2.5.0. The Edge TPU Compiler (Google LLC, 2020f) generates versions of the TensorFlow Lite models compatible with the Edge TPU (Google LLC, 2020d).

Unlike TensorFlow programs, TensorFlow Lite programs perform inferences on individual elements of a dataset, and not on a complete one. Therefore, when performing predictions on a dataset, a loop must be used to iterate over its elements. The third column of results of Tables A.1, A.2 and A.3 shows the image prediction times for Coral Dev Board.

Finally, it is important to note that the first prediction on the Edge TPU is slower than the next ones, since it includes the load of the TensorFlow Lite model into the device memory (Google LLC, 2020a). Thus, after discarding the first prediction, the inference loop starts iterating over the first dataset element, and the prediction is made twice.

In order to perform inferences for GPU using Jetson Nano 2GB EAS, TensorRT 7.1.3.0 has been used. TensorRT is an SDK for high-performance Deep Learning inference that includes an inference optimizer and runtime providing high throughput and low latency for inference applications (NVIDIA Corp., 2021c, 2021d). It comes included as a Debian Package in the image to install in the microSD card (NVIDIA Corp., 2021b).

First, three iPython notebooks have been used to save the OD and OC segmentation and eye fundus classification models using the SavedModel format. Then, the three SavedModel models are exported to ONNX[3] (Open Neural Network Exchange) format using a conversion utility. Finally, the process followed consists of using the Jetson Nano 2GB EAS to generate the CUDA inference engines from the three ONNX models (NVIDIA Corp., 2021c). Then, each inference engine is serialized and saved in a ".plan" file. As a last step, each engine file is read and de-serialized by a program to perform predictions.

When performing inferences with TensorRT, it is possible - and recommendable - to use tensors of various images, since the batch size can have an important effect on the optimizations that TensorRT performs on the models: "Larger batches take longer to process but reduce the average time spent on each sample" (NVIDIA Corp., 2021d).

However, the number of images of the input tensor (i.e., the batch size) is conditioned by the amount of available memory in a Jetson Nano 2GB EAS when generating a CUDA inference engine. Therefore, the compromise solution adopted in this study consists of splitting an input tensor into batches of ten images, to achieve a performance improvement compared to making predictions on individual images.

For each dataset, the image prediction time has been calculated as the mean of the ones for all the batches in which the dataset is split. For each batch, the image prediction time is calculated as the batch prediction time divided by the batch size (i.e., ten images). Thus, for all the datasets used in the experimental tests, the mean prediction time per image and the standard deviation are presented in the fourth and the fifth column of results of Tables A.1, A.2 and A.3.

---

[3] https://onnx.ai

Moreover, as with Cloud GPU, the first prediction when using Jetson Nano 2GB Maxwell GPU is slower than the next ones because of some necessary memory initializations and allocations. Thus, after the first prediction, the inference loop starts iterating over the first batch of the dataset, so that this prediction is made twice but discarded the first time.

Jetson Nano devices are designed to optimize power efficiency (NVIDIA Corp., 2021h). They support two power modes: 5W (5 watts) and MaxN (10 watts). These two modes allow various configurations with several CPU frequencies and numbers of cores online.

For each power mode, NVIDIA provides a predefined configuration by setting the number of CPU cores online and the CPU, GPU, and memory frequencies to preselected values. The current power mode can be toggled between MaxN (mode ID 0) and 5W (mode ID 1) using a specific command or a GUI front end. The default power mode for Jetson Nano 2GB is MaxN.

Finally, it is important to note that the batch size used to perform inferences using Colab notebooks when obtaining the Cloud GPU and TPU image prediction times have also been set to ten images, to make fairer comparisons with the results obtained for Jetson Nano 2GB.

Figs. 9, 10 and 11 show the respective graphical representations from the image prediction times presented in Tables A.1, A.2 and A.3.

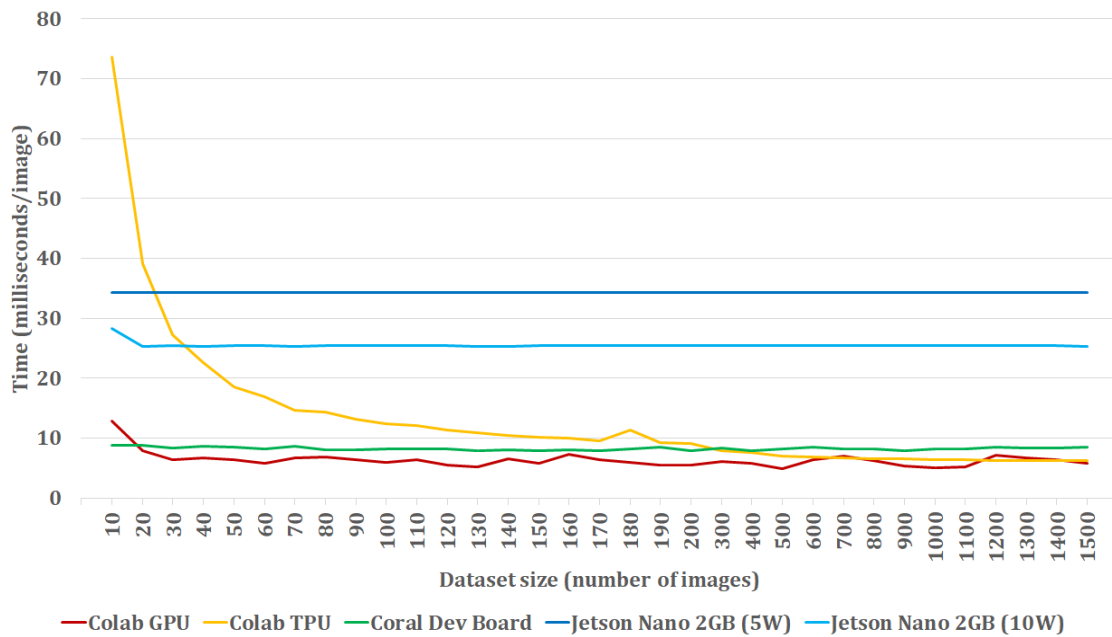

Fig. 9. Image prediction times for segmentation of OD.

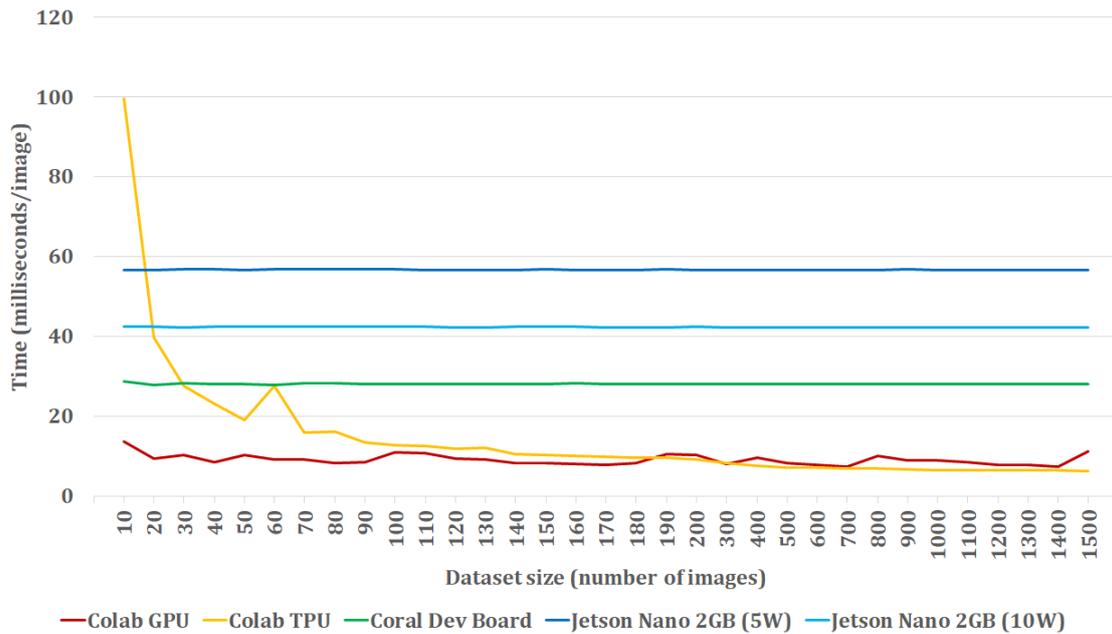

Fig. 10. Image prediction times for segmentation of OC.

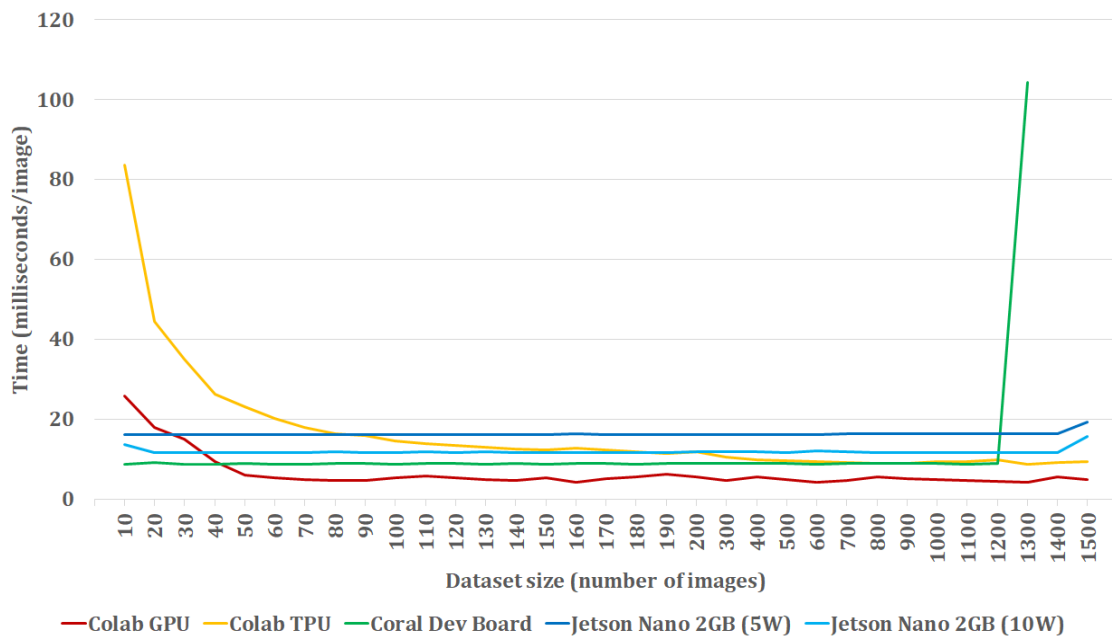

Fig. 11. Image prediction times for eye fundus classification.

Regarding the energy efficiency of the proposed implementations, to obtain the consumption values for the two EAS when performing inferences, an Innovateking-EU TC66C type C USB tester[4] has been connected between the power adapter and the EAS used. The minimum period between measures is one second. Voltage and current values are transmitted in real time to a PC connected to the tester through a micro-USB port. Using the tester program for Windows, these values can be exported to an Excel sheet to be processed.

---

[4] https://t.ly/-DIyA

In order to perform the experimental tests, ten datasets have been selected from those previously used to obtain the image prediction times (shown in Tables A.1, A.2 and A.3). When measuring the current and voltage values for each inference type - OD and OC segmentation, and fundus classification - and each EAS - Coral Dev Board (Edge TPU), and Jetson Nano 2GB (Maxwell GPU) operating in 5W (5 watts) and MAXN (10 watts) power modes - a period of ten seconds before the load of each dataset, and a period of five seconds between a dataset load and the set of predictions have been left in order to identify correctly the power values that effectively correspond to the inferences.

From the Excel sheets generated by the PC program connected to the USB tester, the voltage and current values obtained in each measure, expressed in volts and amperes respectively, have been used to calculate the corresponding power values in watts.

Figs. 12 to 20 show the power values obtained for each inference type and each EAS. It can be observed that, for each of the ten datasets, there is a first set of power values higher than the minimum ones, that correspond to a dataset load, followed by a second set of power values, higher than the first ones, that correspond to the performance of the inferences.

In these figures, it is possible to identify the period of ten seconds before each dataset load, as well as the period of five seconds between a dataset load and the set of predictions. Moreover, in the figures corresponding to predictions performed by Jetson Nano 2GB (Figs. 13, 14, 16, 17, 19 and 20) it can be observed the existence of a first set of power values higher than the minimum ones, that correspond to the load of the CUDA inference engines.

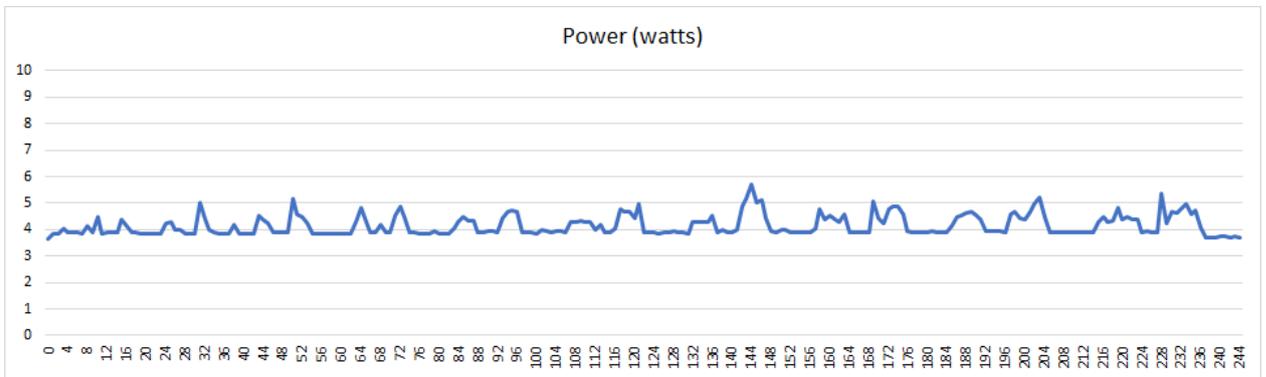

Fig. 12. Power values for optic disc segmentation using Edge TPU.

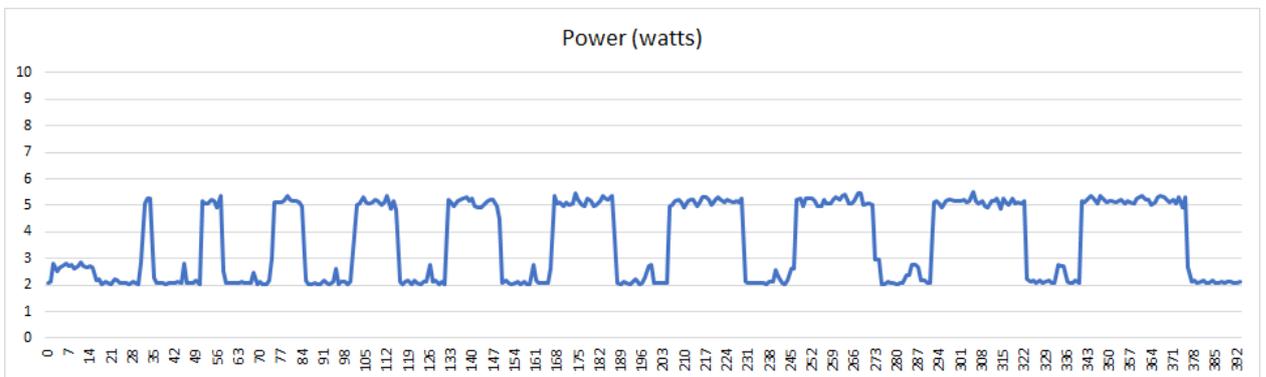

Fig. 13. Power values for optic disc segmentation using Maxwell GPU (5W).

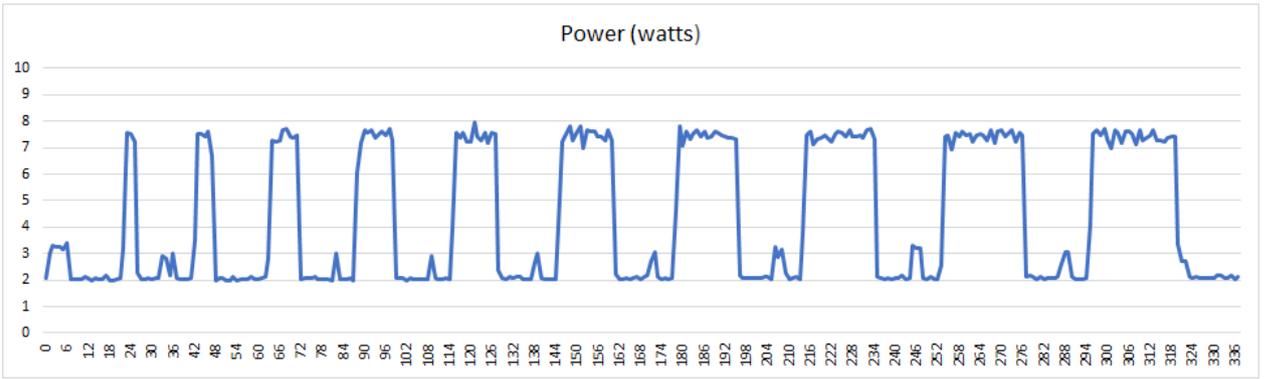
Fig. 14. Power values for optic disc segmentation using Maxwell GPU (10W).

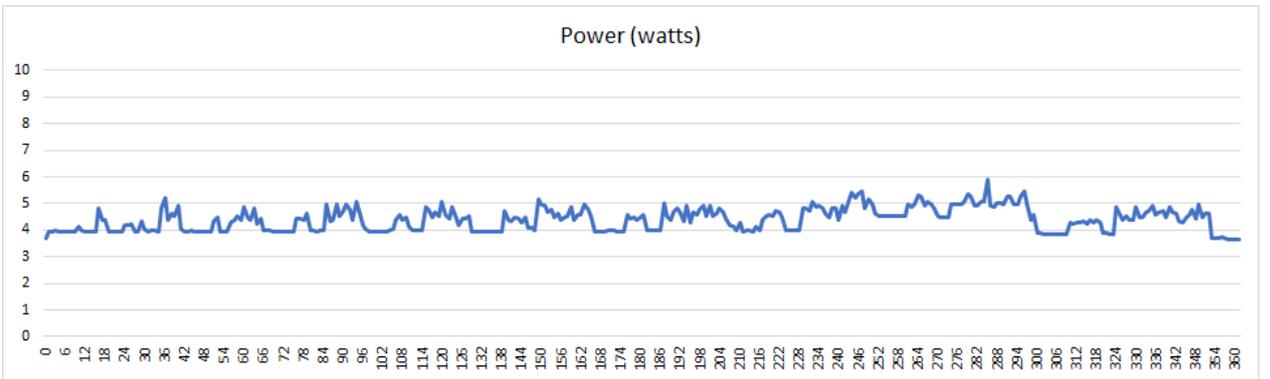
Fig. 15. Power values for optic cup segmentation using Edge TPU.

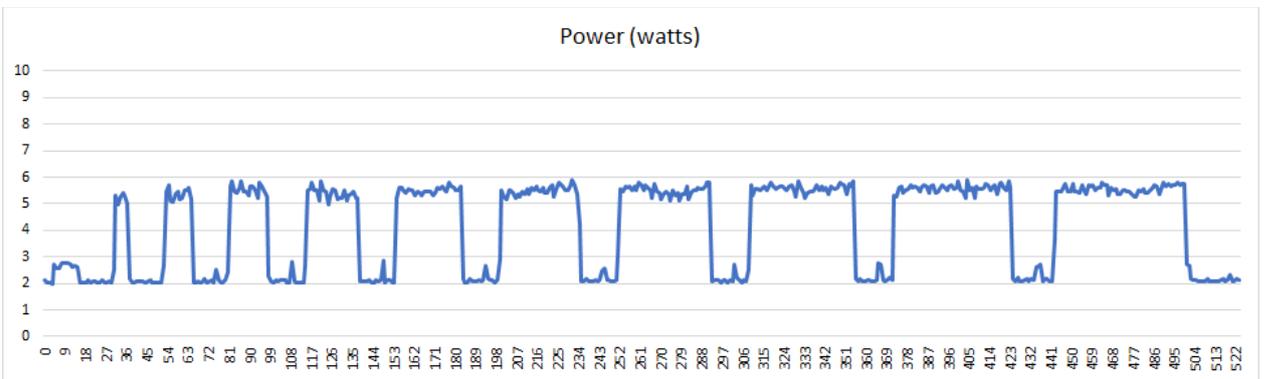
Fig. 16. Power values for optic cup segmentation using Maxwell GPU (5W).

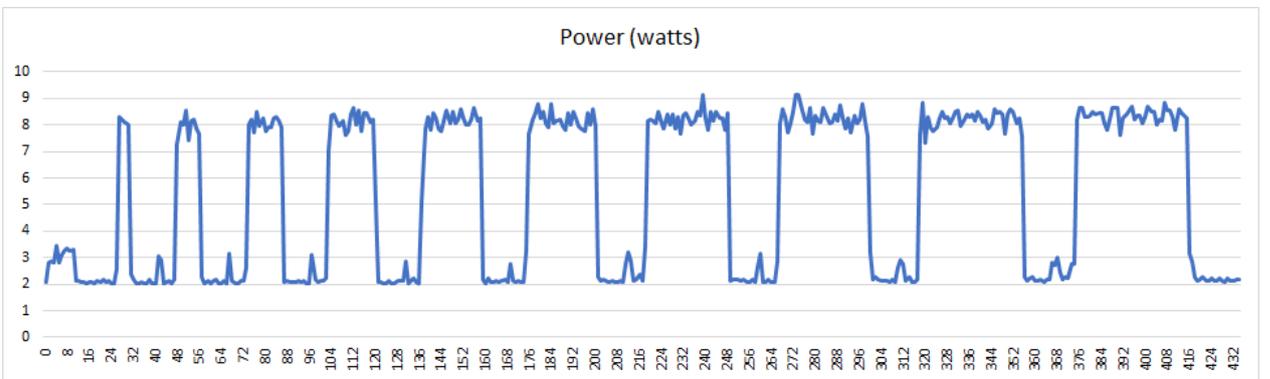
Fig. 17. Power values for optic cup segmentation using Maxwell GPU (10W).

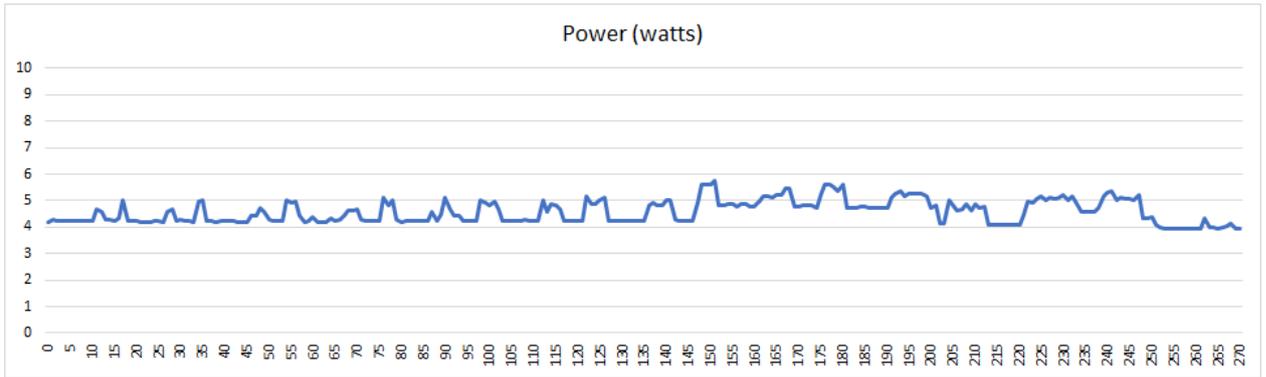
Fig. 18. Power values for eye fundus classification using Edge TPU.

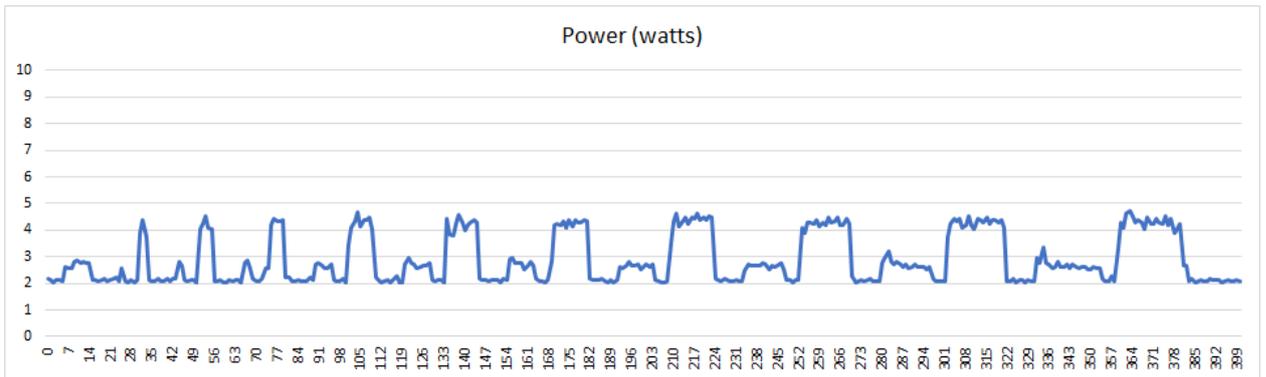
Fig. 19. Power values for eye fundus classification using Maxwell GPU (5W).

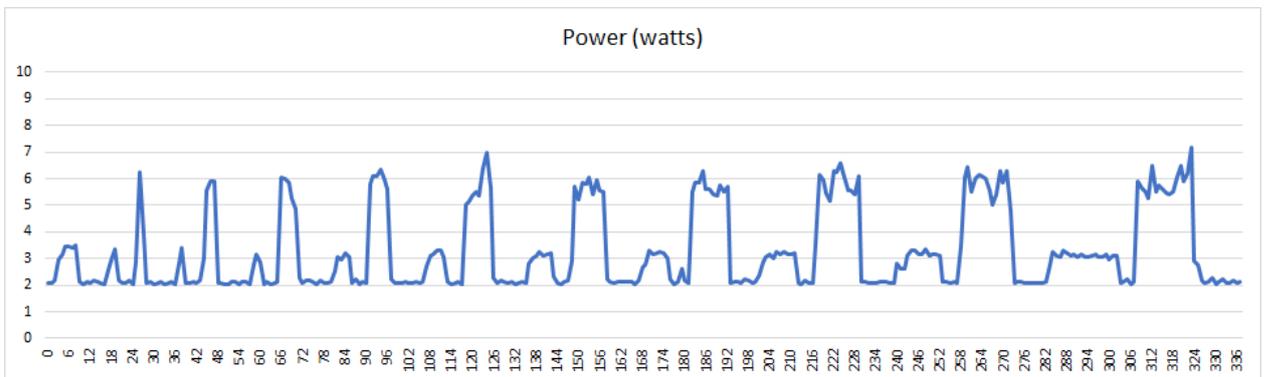
Fig. 20. Power values for eye fundus classification using Maxwell GPU (10W).

Tables B.1, B.2 and B.3 (see Appendix B) show the mean prediction power for each dataset size, EAS and inference type. For each set of inferences, each power value has been calculated as the sum of the products of the current and voltage values measured by the USB tester during the performance of the predictions, divided by the number of measures. Each mean value has been calculated considering only stable power values, thus discarding those obtained at the beginning and at the end of the set of predictions.

The prediction consumption per image for each inference type, EAS and dataset size, has been calculated as the product of the mean power value for the corresponding set of predictions by the image prediction time for the dataset size. These times were previously obtained, and shown in Tables A.1, A.2 and A.3.

# 5. Discussion.

Once it has been proved that predictions made by the two EAS chosen for our study are similar enough to those performed by Google Cloud GPUs and TPUs and, therefore, both devices - Coral Dev Board and Jetson Nano 2GB - are suitable for segmentation and classification of fundus images, it is necessary to assess their performance.

First, for optic disc and cup segmentation (Figs. 9 and 10), image prediction times remain practically stable for Coral Dev Board and Jetson Nano 2GB and could be considered approximately as constant values. However, for eye fundus classification (Fig. 11), image prediction times also remain stable for both EAS until anomalous results are obtained from a dataset size that is specific for each device.

This fact can be explained considering that the image resolution accepted by the classification CNN is 224x224 pixels with three color channels, which is greater than the image resolution accepted by the segmentation U-Nets, consisting of 128x128 pixels with three color channels.

For Coral Dev Board, a very high prediction time is obtained when processing the dataset with 1300 images. Also, a memory error arises when the program tries to load the dataset with 1400 images into the EAS RAM. However, since the memory size of Jetson Nano 2GB is twice the one for Coral Dev Board (i.e., 1 GB), the EAS only provides an anomalous result when processing the dataset with 1500 images. These results are written in bold in Table A.3.

However, it can be expected that more normal classification times for eye fundus datasets with large sizes can be obtained when using higher performance models of both EAS families. In particular, Coral provides a version of its development board with 4GB of RAM, which has the same specifications of the original product but is equipped with the quadruple amount of memory[5]. Also, Jetson Nano Developer Kit[6] has the same specifications for CPU and GPU than Jetson Nano 2GB but doubles the RAM size.

As expected, image prediction times for Jetson Nano 2GB operating in 5W mode are higher than those obtained in MaxN mode for segmentation of OD and OC, and for eye fundus classification. These results are logical considering that the number of CPU cores online as well as the CPU and GPU maximal frequencies are lower in the predefined configuration for 5W power mode than for MaxN (10 watts) power mode (NVIDIA Corp., 2021h).

As for image prediction times using Coral Dev Board, they are lower than the ones obtained for Jetson Nano 2GB in MaxN power mode. This result is also expected, since quantized models for the Edge TPU are smaller and faster (Google, 2021b), and this device is designed to perform extremely fast and power-efficient ML inferencing. An Edge TPU can execute up to four trillion operations per second using only two watts of power[7].

---

[5] https://coral.ai/products
[6] https://www.nvidia.com/en-us/autonomous-machines/jetson-store/
[7] https://coral.ai/docs/edgetpu/faq#what-is-the-edge-tpus-processing-speed

Table 8 shows the minimum speed-ups (SUs) in the three scenarios - optic disc (Table A.1) and cup (Table A.2) segmentation, and eye fundus classification (Table A.3) - for Coral Dev Board (Edge TPU) compared with Jetson Nano 2GB (Maxwell GPU) operating in MaxN power mode (Eq. (1)), and for Jetson Nano 2GB in MaxN mode compared with the same EAS in 5W mode (Eq. (2)). The anomalous results obtained for eye fundus classification have been discarded for the calculation of the speed-ups.

ET PT: Image prediction time for Edge TPU.
MG MaxN PT: Image prediction time for Maxwell GPU in MaxN power mode.
MG 5W PT: Image prediction time for Maxwell GPU in 5W power mode.

$$SU\ ET\ MG\ MaxN = \frac{MG\ MaxN\ PT}{ET\ PT} \qquad (1)$$

$$SU\ MG\ MaxN\ MG\ 5W = \frac{MG\ 5W\ PT}{MG\ MaxN\ PT} \qquad (2)$$

|  | Edge TPU vs. Maxwell GPU MaxN mode | Maxwell GPU MaxN mode vs. 5W mode |
| --- | --- | --- |
| Optic disc | 2.88 | 1.21 |
| Optic cup | 1.47 | 1.33 |
| Eye fundus image | 1.25 | 1.19 |

Table 8. Minimum speed-ups for Edge TPU and Maxwell GPU.

Moreover, in general terms the corresponding results for Google Cloud GPU produce a delimited and relative narrow interval of values, so that they can be considered as stable in the three scenarios: optic disc and cup segmentation, and eye fundus classification (Figs. 9, 10 and 11).

On the other hand, from the obtained results for Cloud TPU prediction times (Tables A.1, A.2 and A.3), their graphical representations as the dataset size (i.e., the number of images) increases can be approximated to a hyperbola plus a constant component (Figs. 9, 10 and 11). Thus, to explain the obtained results, we propose the following equation for Cloud TPU prediction times.

*n*: Dataset size (number of images).
*CT PT* (*n*): Image prediction time for Cloud TPU.
*CT PT* (*OT*): Image prediction time for Cloud TPU (overload term).
*CT PT* (*IT*): Image prediction time for Cloud TPU (independent term).

$$CT\ PT\ (n) = \frac{CT\ PT\ (OT)}{n} + CT\ PT\ (IT) \qquad (3)$$

The function proposed (Eq. (3)) has two summands: The first one is inversely proportional to the dataset size, whereas the second one is independent of this size, and it is the value to which the function tends as the dataset size increases.

For optic disc and cup segmentation, and eye fundus classification, we can observe that for small dataset sizes, image prediction times are far greater for Cloud TPU than for Cloud GPU (Figs. 9, 10 and 11). This result can be explained considering the delays due to the transmission of information through a network (Díaz-del-Río et al., 2016).

It must be noted that the communication between the CPU and the TPU (Google, 2021a) is not local. Consequently, we must consider the existence of a data transmission time, that can be considered as bounded unless a technical incidence occurs. Cloud TPUs are more helpful for training networks[8] than for performing inferences on small datasets. This is logical as they were designed for this purpose.

Finally, when comparing TPU prediction times (Figs. 9, 10 and 11), we can observe that for small datasets, image prediction times are smaller for the Edge TPU compared with those ones for the Cloud TPU.

However, as the size of the dataset (*n*) on which inferences are performed increases, the network data transmission time - which we can identify with the overload term denoted *CT PT* (*OT*) in Eq. (3) - becomes less important compared with the total prediction time for the complete dataset, so that the image prediction time, denoted as *CT PT* (*n*) in Eq. (3), tends to the independent term, denoted *CT PT* (*IT*), which we can be identify with the effective image prediction time by the Cloud TPU.

It is important to note that the Cloud TPU performance is much greater than that of the Edge TPU, since the latter has been designed for model inferencing[8] and not for training of complex and large ML models.

Eq. (4) expresses the ideal speed-up in the Cloud TPU performance versus that of the Edge TPU for a sufficiently large dataset. The term ET PT denotes the Edge TPU image prediction time. Since the Cloud TPU performance is much greater than that of Edge TPU coprocessor, for a sufficiently large dataset size (*n*), the total prediction time using the Cloud TPU plus the data transmission time will end up being smaller than the total time for the inferences on this dataset using the Edge TPU.

$$SU\ CT\ ET = \lim_{n \to \infty} \frac{ET\ PT}{CT\ PT\ (n)} = \lim_{n \to \infty} \frac{ET\ PT}{\frac{CT\ PT\ (OT)}{n} + CT\ PT\ (IT)} = \frac{ET\ PT}{CT\ PT\ (IT)} \quad (4)$$

Summarizing the results obtained in terms of inference times, for the worst case (OC segmentation), inference times are less than 29 ms per image using Coral Dev Board and less than 43 ms per image using Jetson Nano 2GB in MaxN (10 watts) power mode. For the best case, prediction times less than 9 ms per image using Coral Dev Board for OD segmentation, and less than 14 ms per image using Jetson Nano 2GB in MaxN power mode for fundus classification have been achieved. Thus, integrating specific ML hardware on embedded devices allows many complex segmentation and classification tasks to be performed in real time using these devices. In this sense, it would be convenient to integrate ML hardware in medical image acquisition instruments, to enable them to analyze the captured images.

Both devices perform inferences in very reasonable times of a few milliseconds with an acceptable accuracy. However, due to the ML accelerator architecture of each EAS, the Coral Dev Board Edge TPU delivers prediction times per image smaller than those ones provided by the Jetson Nano 2GB Maxwell GPU. Nevertheless, the GPU of this last device allows it to make predictions with a greater accuracy. Therefore, the choice of using one SBC or the other will depend on the nature of the specific problem, in which the accuracy can be more important than inference times or vice versa.

---

[8] https://coral.ai/docs/edgetpu/faq#how-is-the-edge-tpu-different-from-cloud-tpus

In relation to the energy usage of the proposed implementations, from the data presented in Tables B.1, B.2 and B.3 (see Appendix B), a summary table of results (Table 9) has been obtained in order to show the mean power and the image prediction consumption for each type of inference and EAS.

|  | Edge TPU | | Maxwell GPU (5W) | | Maxwell GPU (10W) | |
| --- | --- | --- | --- | --- | --- | --- |
| OD segmentation | $4.6 \pm 0.2$ | $38.0 \pm 2.1$ | $5.2 \pm 0.0$ | $176.8 \pm 1.5$ | $7.5 \pm 0.0$ | $190.0 \pm 1.2$ |
| OC segmentation | $4.7 \pm 0.2$ | $132.1 \pm 4.9$ | $5.5 \pm 0.1$ | $310.9 \pm 6.8$ | $8.2 \pm 0.1$ | $344.6 \pm 3,8$ |
| Fundus classification | $5.1 \pm 0.2$ | $45.2 \pm 2.2$ | $4.3 \pm 0.1$ | $69.3 \pm 1.8$ | $5.9 \pm 0.1$ | $68.6 \pm 1.4$ |

Table 9. Mean powers and image prediction consumptions (in watts and millijoules).

From the obtained results, it can be observed that Coral Dev Board (Edge TPU) is the EAS with the lowest consumption value for OD and OC segmentation, and also for fundus classification, and is also the device that takes less time to perform inferences. This last fact was already known since image prediction times were previously obtained (see Tables A.1, A.2 and A.3).

This very significant energy reduction can be explained considering that the Edge TPU is specifically designed for fast and power-efficient ML inferencing[8]. Moreover, the Edge TPU runs a quantized integer version of the problem, while the Maxwell GPU runs the 32-bit floating point version. We have shown that this quantization has little effect on the problem results.

As for Jetson Nano 2GB (Maxwell GPU), the EAS uses less power when operating in 5W mode than in MAXN mode. Power values for 5W mode do not reach 6 watts for any dataset size when performing OC and OD segmentation, and even less for fundus classification (around 4.3 watts).

However, for the three inference types and all the dataset sizes, image prediction times are higher for 5W mode than for MAXN mode, as shown is Tables A.1, A.2 and A.3. For this reason, consumption values are not very different between both power modes. Moreover, for fundus classification (Table B.3), most of the consumption values are slightly higher when Jetson Nano 2GB operates in 5W mode.

Therefore, when performing resource demanding tasks, operating in 5W mode does not improve the energy usage for Jetson Nano 2GB, at least in our study cases. Even though power values (around 5 watts) are lower for 5W mode, consumption values increase due to the higher image prediction times for that mode. There could be a difference between both power modes in favor of the 5W mode if the EAS were used to perform lighter tasks regarding the use of resources, but that is not our case.

**6. Conclusions.**

This work has presented the benefits of designing portable diagnostic aid systems for use in remote locations without direct access to health services, in order to achieve early diagnosis and alleviate overcrowding in hospital services. To this end, two single-board computers equipped with ML hardware to execute Edge computing based Artificial Intelligence applications, that we call embedded accelerated systems (EAS), have been presented.

More specifically, Coral Dev Board and NVIDIA Jetson Nano 2GB EAS have been used to implement a segmentation and a classification subsystem for eye fundus images and have been tested experimentally. The first EAS incorporates a Google Edge TPU, whereas the second one is equipped with a Maxwell GPU.

The best results obtained show an inference time of less than 9 ms per image (more than 111 Hertz processing speed), achieved by the Edge TPU when performing optic disc segmentation.

Regarding the energy efficiency of these hardware platforms, it can be observed that Coral Dev Board is the EAS with smaller energy usage values per image, compared with those of Jetson Nano 2GB, for the three types of inferences: OD segmentation (around 38 mJ vs. 177 mJ in 5W mode and 190 mJ in 10W mode), OC segmentation (around 132 mJ vs. 311 mJ in 5W mode and 345 mJ in 10W mode), and fundus classification (around 45 mJ vs. 69 mJ in both modes).

It is also interesting to note that the energy usage, at least in this study, does not improve significantly with low power GPU modes, as with Jetson Nano 2GB in 5W mode. We can see that, although the power usage for the 5W mode is reduced by a factor of about 0.7 compared with the MAXN mode, the time to solve the problems - OD and OC segmentation, and fundus classification - is augmented by approximately 1.35. Thus, the energy reduction in the low power mode for these problems is less than 10% in the best case (OC segmentation).

The results presented in this work demonstrate the feasibility of integrating algorithms that include segmentation and classification tasks in specialized hardware devices (such as the ones analyzed in this work), being able to work in real time, with segmentation and classification accuracies comparable to the processing in high-performance devices (such as Google Cloud GPUs and TPUs), and with a significantly lower power consumption. These results support the initial hypothesis that the use of those specialized hardware systems for diagnostic aid tasks can be made with sufficient guarantees in the medical field.

As future work, we intend to continue this study by using other devices of the NVIDIA Jetson series with higher specifications[6] in order to perform alternative experimental tests on the models used in this work, and compare the new results with the already obtained for Jetson Nano 2GB. We also intend to use Coral accelerator devices[5] to implement the proposed models and perform the experimental tests. Such devices are based on an Edge TPU to enable existing systems to perform ML inferencing.

Moreover, Teikari et al. (2019) proposes integrating ML technology in medical devices to perform high quality image acquisition without the intervention of a properly qualified operator and suggests various scenarios where embedded Deep Learning could be used in routine eye examination. For example, patients could be screened in remote areas by a mobile general healthcare practitioner or could be imaged by a technician in a waiting room before an ophthalmologist appointment.

In this vein, considering the possibility of using the TPU integrated in the Google Tensor SoC[9,10] - used by Pixel 6 and Pixel 6 Pro smartphones - for medical image analysis, ophthalmologists could capture patients' eye fundus images and analyze them in their smartphones. In this sense, the models proposed in this work could be implemented on these ML accelerated smartphones to test the performance achieved.

**Credit Roles.**

J.M.R. Corral: Conceptualization, Investigation, Methodology, Software, Supervision, Writing - original draft, Writing - Review & Editing.
J. Civit-Masot: Conceptualization, Data curation, Methodology, Writing - Original Draft, Writing - Review & Editing.
F. Luna-Perejón: Investigation, Software, Visualization.
I. Díaz-Cano: Investigation, Methodology, Validation.
A. Morgado-Estévez: Conceptualization, Resources, Writing - Original Draft.
M. Domínguez-Morales: Methodology, Resources, Visualization, Writing - Original Draft, Writing - Review & Editing.

**Acknowledgments.**

Figures 3 and 4 are provided courtesy of NVIDIA and Coral, respectively.

**References.**

---

[9] https://blog.google/products/pixel/introducing-google-tensor/
[10] https://www.anandtech.com/show/17032/tensor-soc-performance-efficiency

# Annexes

## A. Image prediction times.

| Dataset (shape) | Colab GPU | Colab TPU | Edge TPU | Maxwell GPU (5W) | Maxwell GPU (10W) |
|---|---|---|---|---|---|
| (10, 128, 128, 3) | 12.92 ± 3.41 | 73.61 ± 3.11 | 8.80 ± 1.10 | 34.34 ± 0.00 | 28.32 ± 0.00 |
| (20, 128, 128, 3) | 7.91 ± 0.73 | 39.09 ± 4.02 | 8.75 ± 1.22 | 34.31 ± 0.01 | 25.37 ± 0.01 |
| (30, 128, 128, 3) | 6.44 ± 0.19 | 27.22 ± 0.47 | 8.30 ± 0.96 | 34.32 ± 0.00 | 25.40 ± 0.06 |
| (40, 128, 128, 3) | 6.71 ± 1.49 | 22.61 ± 2.47 | 8.67 ± 1.15 | 34.31 ± 0.03 | 25.34 ± 0.06 |
| (50, 128, 128, 3) | 6.44 ± 0.72 | 18.55 ± 0.29 | 8.54 ± 1.08 | 34.29 ± 0.01 | 25.41 ± 0.10 |
| (60, 128, 128, 3) | 5.75 ± 0.26 | 16.87 ± 1.64 | 8.27 ± 1.12 | 34.30 ± 0.02 | 25.43 ± 0.07 |
| (70, 128, 128, 3) | 6.68 ± 2.03 | 14.72 ± 0.16 | 8.61 ± 1.12 | 34.32 ± 0.08 | 25.31 ± 0.12 |
| (80, 128, 128, 3) | 6.91 ± 1.66 | 14.31 ± 1.15 | 8.00 ± 1.01 | 34.29 ± 0.04 | 25.43 ± 0.06 |
| (90, 128, 128, 3) | 6.38 ± 1.20 | 13.21 ± 0.13 | 8.08 ± 1.07 | 34.29 ± 0.03 | 25.38 ± 0.09 |
| (100, 128, 128, 3) | 6.00 ± 0.83 | 12.35 ± 0.16 | 8.18 ± 1.06 | 34.30 ± 0.02 | 25.40 ± 0.08 |
| (110, 128, 128, 3) | 6.35 ± 0.98 | 12.04 ± 0.83 | 8.15 ± 1.15 | 34.30 ± 0.04 | 25.38 ± 0.08 |
| (120, 128, 128, 3) | 5.56 ± 0.30 | 11.31 ± 0.12 | 8.19 ± 1.16 | 34.29 ± 0.03 | 25.38 ± 0.09 |
| (130, 128, 128, 3) | 5.27 ± 0.24 | 10.97 ± 0.12 | 7.97 ± 1.04 | 34.30 ± 0.04 | 25.37 ± 0.08 |
| (140, 128, 128, 3) | 6.52 ± 2.27 | 10.48 ± 0.09 | 8.03 ± 1.03 | 34.29 ± 0.03 | 25.36 ± 0.08 |
| (150, 128, 128, 3) | 5.81 ± 1.51 | 10.12 ± 0.68 | 7.92 ± 1.00 | 34.29 ± 0.02 | 25.42 ± 0.10 |
| (160, 128, 128, 3) | 7.33 ± 1.47 | 9.95 ± 0.16 | 8.06 ± 1.25 | 34.30 ± 0.04 | 25.40 ± 0.09 |
| (170, 128, 128, 3) | 6.40 ± 1.42 | 9.60 ± 0.15 | 7.95 ± 1.06 | 34.29 ± 0.03 | 25.40 ± 0.09 |
| (180, 128, 128, 3) | 5.94 ± 1.23 | 11.32 ± 1.71 | 8.15 ± 1.09 | 34.32 ± 0.05 | 25.45 ± 0.09 |
| (190, 128, 128, 3) | 5.57 ± 0.96 | 9.20 ± 0.10 | 8.43 ± 1.12 | 34.29 ± 0.02 | 25.44 ± 0.08 |
| (200, 128, 128, 3) | 5.47 ± 0.80 | 9.12 ± 0.53 | 7.94 ± 1.02 | 34.30 ± 0.03 | 25.44 ± 0.08 |
| (300, 128, 128, 3) | 6.03 ± 1.82 | 7.97 ± 0.11 | 8.38 ± 1.06 | 34.31 ± 0.03 | 25.46 ± 0.10 |
| (400, 128, 128, 3) | 5.77 ± 0.93 | 7.54 ± 0.67 | 7.94 ± 1.05 | 34.32 ± 0.05 | 25.46 ± 0.09 |
| (500, 128, 128, 3) | 4.93 ± 0.23 | 7.04 ± 0.05 | 8.13 ± 1.06 | 34.31 ± 0.04 | 25.40 ± 0.09 |
| (600, 128, 128, 3) | 6.35 ± 1.93 | 6.83 ± 0.03 | 8.47 ± 1.06 | 34.32 ± 0.03 | 25.47 ± 0.11 |
| (700, 128, 128, 3) | 6.94 ± 1.08 | 6.73 ± 0.18 | 8.26 ± 1.06 | 34.32 ± 0.04 | 25.39 ± 0.09 |
| (800, 128, 128, 3) | 6.19 ± 0.74 | 6.59 ± 0.03 | 8.24 ± 1.10 | 34.32 ± 0.03 | 25.39 ± 0.09 |
| (900, 128, 128, 3) | 5.41 ± 0.52 | 6.52 ± 0.03 | 7.84 ± 1.01 | 34.32 ± 0.04 | 25.43 ± 0.13 |
| (1000, 128, 128, 3) | 5.07 ± 0.25 | 6.44 ± 0.02 | 8.27 ± 1.11 | 34.32 ± 0.04 | 25.50 ± 0.41 |
| (1100, 128, 128, 3) | 5.25 ± 1.39 | 6.38 ± 0.03 | 8.23 ± 1.06 | 34.31 ± 0.03 | 25.44 ± 0.17 |
| (1200, 128, 128, 3) | 7.08 ± 1.91 | 6.32 ± 0.02 | 8.48 ± 1.09 | 34.31 ± 0.04 | 25.39 ± 0.15 |
| (1300, 128, 128, 3) | 6.70 ± 1.60 | 6.29 ± 0.04 | 8.36 ± 1.08 | 34.32 ± 0.04 | 25.44 ± 0.09 |
| (1400, 128, 128, 3) | 6.38 ± 1.35 | 6.25 ± 0.02 | 8.42 ± 1.08 | 34.32 ± 0.04 | 25.43 ± 0.09 |
| (1500, 128, 128, 3) | 5.85 ± 1.11 | 6.26 ± 0.05 | 8.44 ± 1.09 | 34.31 ± 0.04 | 25.36 ± 0.08 |

Table A.1. Image prediction times for segmentation of OD (in milliseconds).

| Dataset (shape) | Colab GPU | Colab TPU | Edge TPU | Maxwell GPU (5W) | Maxwell GPU (10W) |
| --- | --- | --- | --- | --- | --- |
| (10, 128, 128, 3) | 13.73 ± 3.32 | 99.69 ± 23.03 | 28.66 ± 0.15 | 56.68 ± 0.00 | 42.33 ± 0.00 |
| (20, 128, 128, 3) | 9.43 ± 0.31 | 39.73 ± 1.53 | 27.71 ± 1.29 | 56.71 ± 0.06 | 42.54 ± 0.08 |
| (30, 128, 128, 3) | 10.38 ± 1.54 | 27.51 ± 0.59 | 28.34 ± 0.76 | 56.74 ± 0.09 | 42.31 ± 0.01 |
| (40, 128, 128, 3) | 8.51 ± 0.29 | 23.07 ± 2.12 | 28.03 ± 1.12 | 56.77 ± 0.08 | 42.48 ± 0.21 |
| (50, 128, 128, 3) | 10.29 ± 2.59 | 19.03 ± 0.80 | 28.08 ± 1.08 | 56.72 ± 0.05 | 42.47 ± 0.13 |
| (60, 128, 128, 3) | 9.17 ± 1.65 | 27.52 ± 2.64 | 27.93 ± 1.29 | 56.75 ± 0.05 | 42.46 ± 0.13 |
| (70, 128, 128, 3) | 9.05 ± 0.88 | 15.93 ± 1.36 | 28.18 ± 0.91 | 56.74 ± 0.08 | 42.38 ± 0.13 |
| (80, 128, 128, 3) | 8.17 ± 0.37 | 16.13 ± 3.21 | 28.26 ± 0.91 | 56.74 ± 0.05 | 42.42 ± 0.09 |
| (90, 128, 128, 3) | 8.40 ± 2.11 | 13.53 ± 0.56 | 28.11 ± 1.13 | 56.73 ± 0.08 | 42.39 ± 0.16 |
| (100, 128, 128, 3) | 10.99 ± 2.70 | 12.74 ± 0.24 | 28.04 ± 1.19 | 56.73 ± 0.05 | 42.45 ± 0.18 |
| (110, 128, 128, 3) | 10.70 ± 2.02 | 12.44 ± 1.15 | 28.14 ± 0.99 | 56.72 ± 0.08 | 42.35 ± 0.14 |
| (120, 128, 128, 3) | 9.34 ± 1.68 | 11.76 ± 0.50 | 28.01 ± 1.15 | 56.72 ± 0.06 | 42.29 ± 0.16 |
| (130, 128, 128, 3) | 9.12 ± 1.28 | 12.11 ± 1.50 | 27.94 ± 1.28 | 56.72 ± 0.09 | 42.25 ± 0.16 |
| (140, 128, 128, 3) | 8.30 ± 0.92 | 10.56 ± 0.15 | 27.97 ± 1.24 | 56.72 ± 0.06 | 42.44 ± 0.19 |
| (150, 128, 128, 3) | 8.18 ± 0.63 | 10.33 ± 0.13 | 28.12 ± 1.06 | 56.73 ± 0.10 | 42.37 ± 0.13 |
| (160, 128, 128, 3) | 8.05 ± 0.40 | 10.04 ± 0.12 | 28.16 ± 0.95 | 56.70 ± 0.09 | 42.34 ± 0.19 |
| (170, 128, 128, 3) | 7.71 ± 0.16 | 9.89 ± 0.62 | 28.10 ± 1.10 | 56.71 ± 0.09 | 42.27 ± 0.18 |
| (180, 128, 128, 3) | 8.20 ± 2.09 | 9.64 ± 0.15 | 28.08 ± 1.07 | 56.69 ± 0.06 | 42.31 ± 0.17 |
| (190, 128, 128, 3) | 10.61 ± 3.14 | 9.69 ± 0.59 | 27.98 ± 1.19 | 56.73 ± 0.06 | 42.28 ± 0.15 |
| (200, 128, 128, 3) | 10.27 ± 2.82 | 9.19 ± 0.15 | 28.09 ± 1.08 | 56.70 ± 0.06 | 42.33 ± 0.18 |
| (300, 128, 128, 3) | 7.95 ± 0.68 | 8.24 ± 0.36 | 27.99 ± 1.16 | 56.70 ± 0.08 | 42.23 ± 0.11 |
| (400, 128, 128, 3) | 9.62 ± 2.77 | 7.59 ± 0.12 | 27.98 ± 1.20 | 56.69 ± 0.05 | 42.25 ± 0.18 |
| (500, 128, 128, 3) | 8.30 ± 1.49 | 7.23 ± 0.09 | 28.06 ± 1.17 | 56.71 ± 0.13 | 42.20 ± 0.17 |
| (600, 128, 128, 3) | 7.85 ± 0.72 | 7.11 ± 0.19 | 28.05 ± 1.09 | 56.71 ± 0.07 | 42.24 ± 0.17 |
| (700, 128, 128, 3) | 7.39 ± 0.13 | 6.84 ± 0.05 | 28.01 ± 1.16 | 56.70 ± 0.08 | 42.23 ± 0.15 |
| (800, 128, 128, 3) | 10.12 ± 2.86 | 6.81 ± 0.17 | 27.99 ± 1.20 | 56.71 ± 0.07 | 42.15 ± 0.17 |
| (900, 128, 128, 3) | 9.00 ± 2.07 | 6.69 ± 0.11 | 28.01 ± 1.19 | 56.73 ± 0.08 | 42.16 ± 0.15 |
| (1000, 128, 128, 3) | 8.86 ± 1.54 | 6.54 ± 0.04 | 28.00 ± 1.21 | 56.71 ± 0.07 | 42.16 ± 0.17 |
| (1100, 128, 128, 3) | 8.38 ± 1.09 | 6.55 ± 0.06 | 28.01 ± 1.19 | 56.70 ± 0.08 | 42.25 ± 0.15 |
| (1200, 128, 128, 3) | 7.85 ± 0.66 | 6.41 ± 0.04 | 27.98 ± 1.20 | 56.70 ± 0.07 | 42.18 ± 0.15 |
| (1300, 128, 128, 3) | 7.73 ± 0.36 | 6.42 ± 0.13 | 27.99 ± 1.12 | 56.70 ± 0.07 | 42.20 ± 0.13 |
| (1400, 128, 128, 3) | 7.33 ± 0.13 | 6.34 ± 0.04 | 27.99 ± 1.06 | 56.71 ± 0.07 | 42.20 ± 0.17 |
| (1500, 128, 128, 3) | 11.19 ± 3.23 | 6.33 ± 0.04 | 27.99 ± 1.08 | 56.72 ± 0.08 | 42.29 ± 0.15 |

Table A.2. Image prediction times for segmentation of OC (in milliseconds).

| Dataset (shape) | Colab GPU | Colab TPU | Edge TPU | Maxwell GPU (5W) | Maxwell GPU (10W) |
|---|---|---|---|---|---|
| (10, 224, 224, 3) | 25.84 ± 5.87 | 83.52 ± 2.77 | 8.75 ± 1.24 | 16.19 ± 0.00 | 13.60 ± 0.00 |
| (20, 224, 224, 3) | 17.92 ± 3.40 | 44.38 ± 0.92 | 9.24 ± 1.15 | 16.21 ± 0.01 | 11.55 ± 0.03 |
| (30, 224, 224, 3) | 15.00 ± 2.94 | 34.98 ± 9.18 | 8.67 ± 1.42 | 16.22 ± 0.02 | 11.53 ± 0.01 |
| (40, 224, 224, 3) | 9.33 ± 2.84 | 26.17 ± 0.66 | 8.77 ± 1.28 | 16.20 ± 0.02 | 11.55 ± 0.03 |
| (50, 224, 224, 3) | 5.92 ± 1.11 | 23.01 ± 3.73 | 8.90 ± 1.36 | 16.21 ± 0.02 | 11.57 ± 0.04 |
| (60, 224, 224, 3) | 5.31 ± 0.62 | 20.09 ± 1.82 | 8.67 ± 1.35 | 16.21 ± 0.02 | 11.55 ± 0.04 |
| (70, 224, 224, 3) | 4.90 ± 0.37 | 17.91 ± 0.41 | 8.77 ± 1.34 | 16.21 ± 0.02 | 11.73 ± 0.52 |
| (80, 224, 224, 3) | 4.60 ± 0.17 | 16.40 ± 0.25 | 8.86 ± 1.33 | 16.22 ± 0.03 | 11.95 ± 0.67 |
| (90, 128, 128, 3) | 4.76 ± 0.97 | 15.84 ± 1.38 | 9.02 ± 1.39 | 16.20 ± 0.04 | 11.54 ± 0.03 |
| (100, 224, 224, 3) | 5.43 ± 1.25 | 14.51 ± 0.24 | 8.79 ± 1.37 | 16.19 ± 0.02 | 11.52 ± 0.04 |
| (110, 224, 224, 3) | 5.72 ± 0.92 | 13.80 ± 0.16 | 8.99 ± 1.34 | 16.21 ± 0.04 | 11.86 ± 0.75 |
| (120, 224, 224, 3) | 5.38 ± 0.72 | 13.49 ± 0.28 | 8.84 ± 1.32 | 16.21 ± 0.04 | 11.52 ± 0.03 |
| (130, 224, 224, 3) | 4.95 ± 0.58 | 13.03 ± 0.22 | 8.74 ± 1.35 | 16.20 ± 0.04 | 11.88 ± 0.39 |
| (140, 224, 224, 3) | 4.57 ± 0.37 | 12.55 ± 0.22 | 8.91 ± 1.36 | 16.20 ± 0.03 | 11.53 ± 0.02 |
| (150, 224, 224, 3) | 5.30 ± 1.39 | 12.37 ± 0.20 | 8.77 ± 1.32 | 16.21 ± 0.05 | 11.55 ± 0.09 |
| (160, 224, 224, 3) | 4.30 ± 0.12 | 12.64 ± 0.66 | 8.82 ± 1.32 | 16.24 ± 0.09 | 11.53 ± 0.03 |
| (170, 224, 224, 3) | 5.13 ± 1.43 | 12.36 ± 1.00 | 8.84 ± 1.36 | 16.20 ± 0.03 | 11.53 ± 0.04 |
| (180, 224, 224, 3) | 5.54 ± 1.62 | 11.75 ± 0.20 | 8.70 ± 1.34 | 16.22 ± 0.04 | 11.55 ± 0.09 |
| (190, 224, 224, 3) | 6.26 ± 1.35 | 11.50 ± 0.17 | 9.02 ± 1.37 | 16.21 ± 0.03 | 11.53 ± 0.02 |
| (200, 224, 224, 3) | 5.59 ± 1.17 | 11.80 ± 0.81 | 8.86 ± 1.33 | 16.21 ± 0.03 | 11.95 ± 0.89 |
| (300, 224, 224, 3) | 4.59 ± 0.35 | 10.43 ± 0.19 | 8.95 ± 1.35 | 16.23 ± 0.03 | 11.74 ± 0.53 |
| (400, 224, 224, 3) | 5.47 ± 1.37 | 9.74 ± 0.06 | 8.88 ± 1.32 | 16.23 ± 0.04 | 11.75 ± 0.50 |
| (500, 224, 224, 3) | 4.85 ± 0.72 | 9.63 ± 0.38 | 8.94 ± 1.34 | 16.23 ± 0.06 | 11.53 ± 0.03 |
| (600, 224, 224, 3) | 4.22 ± 0.29 | 9.32 ± 0.09 | 8.78 ± 1.33 | 16.23 ± 0.03 | 12.08 ± 2.44 |
| (700, 224, 224, 3) | 4.73 ± 1.55 | 9.05 ± 0.08 | 8.95 ± 1.34 | 16.25 ± 0.06 | 11.88 ± 1.73 |
| (800, 224, 224, 3) | 5.66 ± 1.35 | 8.99 ± 0.11 | 8.92 ± 1.33 | 16.25 ± 0.07 | 11.56 ± 0.10 |
| (900, 224, 224, 3) | 5.04 ± 1.07 | 8.98 ± 0.06 | 8.87 ± 1.33 | 16.24 ± 0.06 | 11.58 ± 0.04 |
| (1000, 224, 224, 3) | 4.91 ± 0.66 | 9.39 ± 0.29 | 8.82 ± 1.34 | 16.26 ± 0.07 | 11.60 ± 0.05 |
| (1100, 224, 224, 3) | 4.54 ± 0.50 | 9.40 ± 0.07 | 8.79 ± 1.33 | 16.26 ± 0.07 | 11.59 ± 0.06 |
| (1200, 224, 224, 3) | 4.39 ± 0.25 | 9.72 ± 0.24 | 8.91 ± 1.32 | 16.26 ± 0.07 | 11.63 ± 0.11 |
| (1300, 224, 224, 3) | 4.19 ± 0.30 | 8.64 ± 0.11 | **104.4 ± 41.01** | 16.28 ± 0.11 | 11.66 ± 0.32 |
| (1400, 224, 224, 3) | 5.58 ± 1.71 | 9.06 ± 0.19 | **memory error** | 16.30 ± 0.12 | 11.63 ± 0.08 |
| (1500, 224, 224, 3) | 4.77 ± 1.32 | 9.40 ± 0.06 | **memory error** | **19.24 ± 29.75** | **15.62 ± 26.59** |

Table A.3. Image prediction times for eye fundus classification (in milliseconds).

**Notes.**

The first two columns of results of Tables A.1 and A.2 are the prediction times per image obtained using the Colaboratory iPython notebook development environment for segmentation of OD and OC, respectively, whereas the first two columns of results of Table A.3 show the prediction times per image for fundus classification. Thus, the mean prediction time per image and the standard deviation are shown in the two first columns of results of Tables A.1, A.2 and A.3.

As for Coral Dev Board, the prediction times using the Edge TPU for optic disc and cup segmentation, and for eye fundus classification are presented in the third column of results of Tables A.1, A.2 and A.3, respectively.

Regarding GPU inferences using Jetson Nano 2GB EAS, the mean prediction time per image and the standard deviation are presented in the fourth and the fifth column of results - 5W power mode and MaxN power mode (10W) - of Tables A.1, A.2 and A.3.

## B. Mean power and image prediction energy use.

| Dataset (shape) | Edge TPU | | Maxwell GPU (5W) | | Maxwell GPU (10W) | |
|---|---|---|---|---|---|---|
| (100, 128, 128, 3) | 4.2 ± 0.1 | 34 | 5.2 ± 0.1 | 178 | 7.4±0.2 | 188 |
| (200, 128, 128, 3) | 4.5 ± 0.4 | 36 | 5.2 ± 0.1 | 178 | 7.5±0.1 | 191 |
| (300, 128, 128, 3) | 4.6 ± 0.3 | 39 | 5.1 ± 0.1 | 175 | 7.4±0.2 | 188 |
| (400, 128, 128, 3) | 4.6 ± 0.2 | 37 | 5.1 ± 0.1 | 175 | 7.5±0.2 | 191 |
| (500, 128, 128, 3) | 4.6 ± 0.1 | 37 | 5.1 ± 0.1 | 175 | 7.5±0.2 | 191 |
| (600, 128, 128, 3) | 4.7 ± 0.2 | 40 | 5.2 ± 0.1 | 178 | 7.5±0.2 | 191 |
| (700, 128, 128, 3) | 5.1 ± 0.4 | 42 | 5.2 ± 0.1 | 178 | 7.5±0.2 | 190 |
| (800, 128, 128, 3) | 4.7 ± 0.3 | 39 | 5.2 ± 0.1 | 178 | 7.5±0.1 | 190 |
| (900, 128, 128, 3) | 4.7 ± 0.3 | 37 | 5.1 ± 0.1 | 175 | 7.5±0.2 | 191 |
| (1000, 128, 128, 3) | 4.7 ± 0.4 | 39 | 5.2 ± 0.1 | 178 | 7.4±0.2 | 189 |

Table B.1. Mean power and image prediction energy use for optic disc segmentation (in watts and millijoules).

| Dataset (shape) | Edge TPU | | Maxwell GPU (5W) | | Maxwell GPU (10W) | |
|---|---|---|---|---|---|---|
| (100, 128, 128, 3) | 4.5 ± 0.2 | 126 | 5.2 ± 0.2 | 295 | 8.2 ± 0.1 | 348 |
| (200, 128, 128, 3) | 4.8 ± 0.3 | 135 | 5.4 ± 0.2 | 306 | 8.0 ± 0.3 | 339 |
| (300, 128, 128, 3) | 4.5 ± 0.2 | 126 | 5.5 ± 0.2 | 312 | 8.1 ± 0.2 | 342 |
| (400, 128, 128, 3) | 4.7 ± 0.3 | 132 | 5.4 ± 0.2 | 306 | 8.2 ± 0.3 | 346 |
| (500, 128, 128, 3) | 4.6 ± 0.2 | 129 | 5.5 ± 0.1 | 312 | 8.2 ± 0.3 | 346 |
| (600, 128, 128, 3) | 4.7 ± 0.2 | 132 | 5.5 ± 0.2 | 312 | 8.2 ± 0.3 | 346 |
| (700, 128, 128, 3) | 4.7 ± 0.2 | 132 | 5.5 ± 0.2 | 312 | 8.2 ± 0.3 | 346 |
| (800, 128, 128, 3) | 4.9 ± 0.3 | 137 | 5.6 ± 0.1 | 318 | 8.3 ± 0.4 | 350 |
| (900, 128, 128, 3) | 5.1 ± 0.2 | 143 | 5.6 ± 0.1 | 318 | 8.2 ± 0.3 | 346 |
| (1000, 128, 128, 3) | 4.6 ± 0.2 | 129 | 5.6 ± 0.1 | 318 | 8.0 ± 0.3 | 337 |

Table B.2. Mean power and image prediction energy use for optic cup segmentation (in watts and millijoules).

| Dataset (shape) | Edge TPU | | Maxwell GPU (5W) | | Maxwell GPU (10W) | |
|---|---|---|---|---|---|---|
| (100, 224, 224, 3) | 5.0 ± 0.0 | 44 | 4.0 ± 0.3 | 65 | 6.2 ± 0.0 | 71 |
| (200, 224, 224, 3) | 5.0 ± 0.0 | 44 | 4.2 ± 0.2 | 68 | 5.8 ± 0.2 | 69 |
| (300, 224, 224, 3) | 5.0 ± 0.1 | 45 | 4.3 ± 0.1 | 70 | 5.8 ± 0.3 | 68 |
| (400, 224, 224, 3) | 5.0 ± 0.1 | 44 | 4.3 ± 0.2 | 70 | 6.0 ± 0.2 | 71 |
| (500, 224, 224, 3) | 4.9 ± 0.1 | 44 | 4.2 ± 0.2 | 68 | 5.7 ± 0.6 | 66 |
| (600, 224, 224, 3) | 5.0 ± 0.1 | 44 | 4.3 ± 0.1 | 70 | 5.7 ± 0.3 | 69 |
| (700, 224, 224, 3) | 5.6 ± 0.1 | 50 | 4.4 ± 0.1 | 72 | 5.7 ± 0.2 | 68 |
| (800, 224, 224, 3) | 5.5 ± 0.2 | 49 | 4.3 ± 0.1 | 70 | 5.9 ± 0.4 | 68 |
| (900, 224, 224, 3) | 4.8 ± 0.1 | 43 | 4.3 ± 0.1 | 70 | 5.9 ± 0.4 | 68 |
| (1000, 224, 224, 3) | 5.1 ± 0.1 | 45 | 4.3 ± 0.2 | 70 | 5.9 ± 0.5 | 68 |

Table B.3. Mean power and image prediction energy use for eye fundus classification (in watts and millijoules).

**C. Abbreviations.**

| Abbreviation | Description |
|---|---|
| AI | Artificial Intelligence |
| CAD | Computer-aided diagnosis |
| CDR | Cup-to-disc-ratio |
| CNN | Convolutional Neural Network |
| CT | Computerized Tomography |
| DFU | Diabetic Foot Ulcer |
| EAS | Embedded accelerated system |
| FCN | Fully convolutional network |
| ISNT | Inferior, superior, nasal and temporal |
| ML | Machine Learning |
| MRI | Magnetic Resonance imaging |
| MTP | Million trainable parameters |
| MXU | Matrix Multiply Unit |
| OD | Optic disc |
| OC | Optic cup |
| TPU | Tensor Processing Unit |
| SBC | Single-board computer |
| SoC | System on chip |
| SoM | System-on-module |
| ONNX | Open Neural Network Exchange |
| WHO | World Health Organization |

Table C.1. Abbreviations.